\documentclass[12pt]{article}

\usepackage[margin=0.8in]{geometry}
\usepackage{amsmath}
\usepackage{amssymb}
\usepackage[hidelinks]{hyperref}

\usepackage{graphicx}
\usepackage[linesnumbered,algoruled,boxed,lined]{algorithm2e}
\usepackage[noend]{algpseudocode}
\usepackage{url}
\usepackage{balance}  % for  \balance command ON LAST PAGE  (only there!)
\usepackage{enumitem}
\usepackage{comment} %for commenting out whole sections 

\usepackage[belowskip=-9pt,aboveskip=2pt]{caption}
\usepackage{xcolor}%To have color in fonts

\newcommand{\nosemic}{\renewcommand{\@endalgocfline}{\relax}}% Drop semi-colon ;
\newcommand{\dosemic}{\renewcommand{\@endalgocfline}{\algocf@endline}}% Reinstate semi-colon ;
\newcommand{\pushline}{\Indp}% Indent
\newcommand{\popline}{\Indm}% Undent
\let\oldnl\nl% Store \nl in \oldnl
\newcommand{\nonl}{\renewcommand{\nl}{\let\nl\oldnl}}% Remove line number for one line
\begin{document}

%SWARM: Spatial Workload-aware Adaptive Routing Mechanism
%SWARM: Spatial Workload-aware Adaptive Routing Management
%SWARM: Spatial Workload-aware Adaptive Routing Manager
\title{SWARM: Adaptive Load Balancing in Distributed Streaming Systems for Big Spatial Data}
\date{\vspace{-5ex}}

\author{Anas Daghistani, Walid G. Aref, Arif Ghafoor, and Ahmed R. Mahmood\\ 
Purdue University, West Lafayette, IN\\
\{anas, aref, ghafoor, amahmoo\}@purdue.edu}

\maketitle
\begin{abstract}
The proliferation of GPS-enabled devices has led to the development of numerous location-based services. These services need to process massive amounts of spatial data in real-time. The current scale of spatial data cannot be handled using centralized systems. This has led to the development of distributed spatial streaming systems. Existing systems are using static spatial partitioning to distribute the workload. In contrast, the real-time streamed spatial data follows non-uniform spatial distributions that are continuously changing over time. Distributed spatial streaming systems need to react to the changes in the distribution of spatial data and queries. This paper introduces SWARM, a light-weight adaptivity protocol that continuously monitors the  data and query workloads across the distributed processes of the spatial data streaming system, and redistribute and rebalance the workloads soon as performance bottlenecks get detected. SWARM is able to handle multiple query-execution and data-persistence models. A distributed streaming system can directly use SWARM to adaptively rebalance the system's workload among its machines with minimal changes to the original code of the underlying spatial application. Extensive experimental evaluation using real and synthetic datasets illustrate that, on average, SWARM achieves 200\% improvement over a static grid partitioning that is determined based on observing a limited history of the data and query workloads. Moreover, SWARM reduces execution latency on average 4x compared with the other technique. 
\end{abstract}

\sloppy
\section{Introduction} 
The recent growth in spatial data has been phenomenal due to the proliferation of GPS-enabled devices, e.g., smartphones, smart watches, health monitors, and connected vehicles. Also, social networks generate huge deluge of spatial data, e.g., 500 million tweets are created daily, and they can be geotagged~\cite{internetStats}. This growth leads to the development of location-based services, e.g., Internet search engines that return results based on user location, self-driving cars, video games (e.g., Pokemon GO), and ride-sharing services. Five billion Google search queries are generated every day~\cite{internetStats}. Supporting these services place a huge demand on developing real-time, efficient, and scalable systems for processing location-based queries. Therefore, there is a growing demand to develop new systems that are optimized to process big spatial data instead of using general-purpose systems that are not tunable for the needs of spatial data~\cite{mokbel2016thinkingSpatial}.

Distributed data streaming systems have the potential to provide real-time scalable solutions. There is an increasing number of spatial applications that are being implemented using these systems. Examples include Storm~\cite{toshniwal2014storm}, Twitter Heron~\cite{kulkarni2015twitter}, and SparkStreaming~\cite{zaharia2012discretized}. Spatial applications require extending the capabilities of general distributed data streaming systems to support spatial operations and spatial query processing. In particular, spatial partitioning and indexing techniques are needed to support efficient processing of spatial data~\cite{abdelhamid2016cruncher,choi2015processing,lee2015distributed,song2016parallel,yu2015scalable,wu2012parallelizing,mahmood2015tornado,mahmood2018adaptive,chen2017distributed}. 

\begin{figure}
\centering
\includegraphics[width=0.9\textwidth]{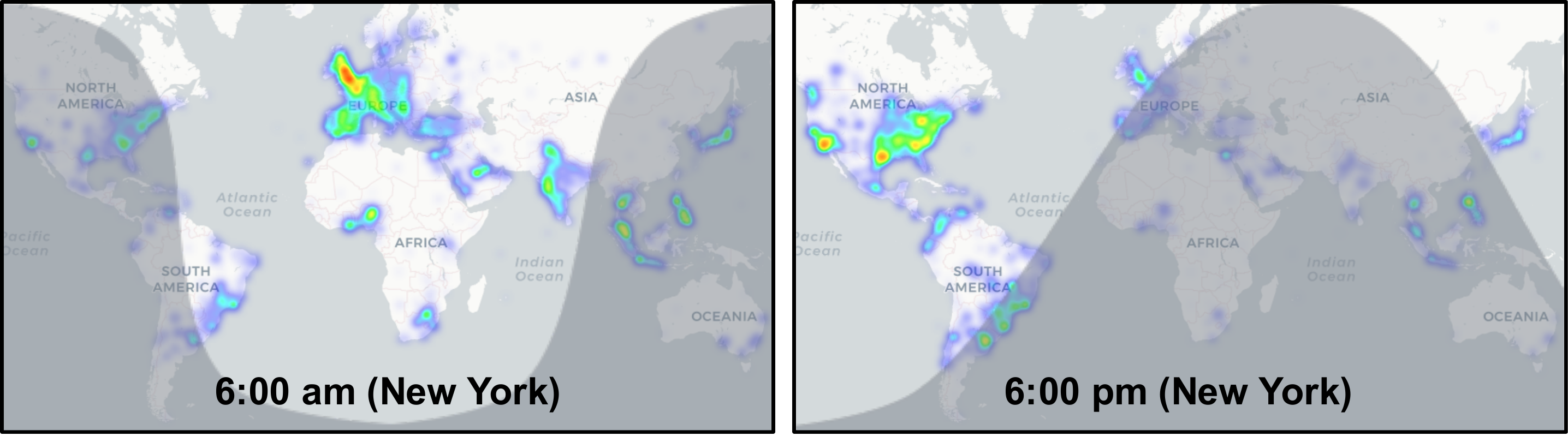}
\caption{\label{fig:tweetsHeatmap} Heatmap of tweets during different times}
\end{figure}

\noindent\textbf{Motivation} A key challenge to improve the performance of a distributed system is to ensure workload balancing across its machines. However, the workload can change rapidly in spatial data applications. The challenge in load balancing stems from the fact that the spatial distributions of data and queries are skewed, and this skewness changes with time and with users' interests. For example, different time-zones can lead to significant changes in the spatial distribution of the data being generated throughout the day. In addition, a major event in a specific location can also lead to the generation of new data and queries pertaining to that event. Figure~\ref{fig:tweetsHeatmap} illustrates heatmaps of tweets generated during one hour at various times of the day. Notice that at 6AM (EDT), Europe and Asia are more active than the Americas. The opposite happens at 6PM (EDT). Moreover, events related to sports, politics, natural disaster, etc., can cause a huge change in the distribution of tweets and queries generated as a large number of users get interested in these events.

\noindent\textbf{Difference from Previous Work. } Most of the existing cluster-based data streaming systems use static data-partitioning schemes to distribute the workload among machines. For this purpose, a limited history of the collected data is used. However, static data-partitioning schemes are not effective in spatial applications due to the rapid changes in spatial data and query distributions as mentioned above. Existing systems, e.g.,~\cite{Aly2015aqwa,Aly2016kangaroo} address this issue by using adaptive mechanisms to update the data-partitioning plan as the workload changes. These systems use a centralized approach to keep statistics and decisions about changing the plan. Deploying a new partitioning plan requires temporary halting the query processor until repartitioning takes place. However, the solutions in~\cite{Aly2015aqwa,Aly2016kangaroo} are not viable for distributed data streaming systems because data and statistics are distributed on different machines. In addition, streams cannot be stopped until repartitioning takes place, and processing should happen in real-time. Other existing adaptive data streaming systems~\cite{mahmood2018adaptive,mahmood2015tornado,chen2017distributed} are designed to work only with spatio-textual data streams and continuous spatial-keyword filter queries. In addition, they only support spatio-textual publish/subscribe applications that result in removing data points as soon as they are processed.

\noindent\textbf{Our Goal. } We propose SWARM, a \textbf{S}patial \textbf{W}orkload-aware \textbf{A}daptive \textbf{R}outing \textbf{M}anager. SWARM is a layer that can be integrated into any distributed data streaming system that processes spatial data. SWARM adaptively load balances the workload among the available machines. SWARM is generic and does not depend on a specific spatial application. It can be used directly with minimal changes to the code of the spatial application. SWARM achieves high machine utilization. This leads to high performance, low response time, and the handling of larger volumes of spatial data and queries. 
\begin{sloppypar}
\noindent\textbf{Challenges and Contributions.} SWARM addresses the following challenges:\\
\textbf{(1)~Variability of optimal partitioning plan due to changes in workload distribution:} The rapid changes in the distribution of spatial data and queries directly affect the optimality in the way the data is partitioned. SWARM adaptively and incrementally alters the data partitions according to the workload changes to achieve higher throughput. SWARM uses a cost model to adaptively change the data partitions to achieve an optimized partitioning.\\
\textbf{(2)~Limited network bandwidth:} High arrival rates of data and queries are expected in applications of distributed data streaming systems. Therefore, the underlying network of these systems can easily become a bottleneck. To overcome this challenge, SWARM adopts a decentralized approach. In particular, SWARM minimizes the load-balancing communication among the machines by maximizing local decision-making 
while using locally collected statistics. \\
\textbf{(3)~Absence of a global system workload state:} 
It is infeasible to collect and exchange statistics with a centralized unit because it introduces high network overhead. SWARM collects and maintains statistics locally in each machine to reduce the communication overhead.\\
\textbf{(4)~Data Repartitioning Overhead:} Data repartitioning can cause communication and computational overhead especially if data is repartitioned unnecessarily. SWARM avoids unnecessary repartitioning of the data by using a probabilistic cost model that predicts the workload of each machine. Upon repartitioning, SWARM moves only the queries without moving the data to avoid the communication overhead. SWARM keeps track of the old data location until the data expires. SWARM avoids unnecessary computations by breaking the process of repartitioning into stages. The amount of computation increases with each stage while the number of involved machines decreases significantly. The last stage involves only the highest overloaded machine that uses efficient heuristic algorithms to reduce its workload.\\
\textbf{(5)~Integrity of the system while avoiding halts during load balancing:} Streams cannot be stopped, and should be processed in real-time. SWARM uses an asynchronous approach for data repartitioning that does not use barriers. SWARM updates its partitions and index using a latch-free~\cite{horikawa2013latch,herlihy2003obstruction,herlihy1991wait,herlihy1990methodology} mechanism.
SWARM does not stop receiving and processing new data during the process of load balancing. SWARM's mechanisms ensure that no results are lost nor are they reported twice. In addition, SWARM knows the location of the data even if the partitions are moved to other machines. If old data is needed to answer a query, SWARM consults the older responsible machines to answer the query.
\end{sloppypar}

\noindent\textbf{Our Approach.} SWARM is a protocol for adaptive load balancing of the machines in a distributed data streaming system that process big spatial data. SWARM is composed of two layers: 1)~the {\em routing layer} that consists of duplicate machines that receive new data points and queries from stream sources, and route them to the appropriate executor machines, and 2)~the {\em execution load-balancing layer} that is composed of load-balancing units that are installed above the original system's executor machines. SWARM uses a probabilistic cost model that predicts the workload of each machine based on changes of the workload's spatial distribution. Therefore, workload reduction is only considered for machines that have \textit{hotspots}. \textit{A hotspot} is a region with a large amount of queries and high probability to receive a lot of new data and queries. 

SWARM does not require prior knowledge about the distribution of data or the queries. SWARM uses an efficient way for locally maintaining statistics in each machine. This makes it possible to decentralize load balancing, maximize local decision making, and reduce the communication overhead. Load balancing is achieved by lazy workload reduction, where only the machine with the highest cost is considered for workload reduction in each round of load balancing. Moreover, the machine, say $m_H$, with the highest cost is allowed to reduce its workload by moving part of its query processing load only to the machine, say $m_L$, with the lowest cost. The reduction of $m_H$'s workload can happen by either 1)~$m_H$ finding a subset of its partitions to move to $m_L$, or 2)~$m_H$ splitting one of its partitions and moving one of the sub-partitions to $m_L$. Efficient algorithms for finding the best subset or split is presented and implemented in SWARM.

The rest of this paper proceeds as follows. Section~\ref{sec:preliminaries} describes the supported data and query models and the notations used in the paper. Section~\ref{sec:costModel} presents a cost model for load balancing. Section~\ref{sec:SWARM} explains SWARM, its indexes, its maintained statistics, and its adaptive load balancing process. Section~\ref{sec:integrity} describes how SWARM preserves system integrity. Section~\ref{sec:experiments} studies the performance of SWARM. Section~\ref{sec:relatedWork} discusses the related work. Section~\ref{sec:conclusions} concludes the paper.

\section{Data and Query Models}\label{sec:preliminaries}
%This section presents the data and query models that are supported by SWARM and the notations that are used throughout the paper. 
SWARM works with any distributed streaming system that processes spatial data using a data processing pipeline. SWARM is designed for tuple-at-a-time systems (e.g., Apache Storm~\cite{toshniwal2014storm}) that target milliseconds latency and not for micro-batched systems (e.g., Spark Streaming~\cite{zaharia2012discretized}) that target sub-second latency.The data and the query streams are redirected to SWARM first. To stress the performance of the system, we assume that the maximum arrival rate of the data stream is higher than the processing capability of the system. This means that the application is trying to fully utilize its machines to process as much data as possible. The main requirement for the data points is to have geo-locations, e.g., Twitter is a good source of geo-tagged tweets that are generated every second. SWARM supports snapshot and continuous queries. A continuous query progressively reports the query results, mainly the data points that satisfy the query's spatial range and its other predicates. Some applications are interested in the recent portion of the data, e.g., the most recent hour. This interest can be expressed as a sliding or a tumbling window. Data expires once it exits the window. SWARM will need to update its statistics accordingly.

As a proof-of-concept, we realize SWARM to adaptively load-balance a location-aware publish-subscribe system. The input stream is geotagged tweets from Twitter, where users subscribe to get tweets in a specific spatial range. A tweet is geo-tagged as a point, say $d$, in space, where $s$ qualifies for a user's subscription, say  Query $q$, if $d$ lies inside $q$'s spatial range. Typically, before SWARM, each query gets replicated into all executor machines, e.g., in an R*-Tree~\cite{beckmann1990rtree}. Each point is directed to only one executor machine, and is checked against all queries using the replicated R*-Tree.

We use the term {\em Hotspot} to refer to a spatial region that receives a large amount of queries and/or data points that is likely to persist for some duration of time. This definition of a \textit{hotspots} excludes spikes of heavy workloads that do not persist for a significant duration of time. 

\section{The Cost Model}\label{sec:costModel}
SWARM distributes the system workload using a cost model that depends on the amount of data points in each partition. The cost model gives higher weight to partitions having a high number of queries. Moreover, the cost model predicts the future workload of each partition based on the workload history. This prediction serves as a scale factor for the overall cost and workload of each partition. Assume that we have a distributed streaming system, say $S$, that has a set $M$ of executor machines. Each machine $m \in M$ holds some partitions $P_m$, where $|P_m|=n_m$, $n_m$ is the number of partitions in Machine $m$. Each partition $p \in P_m$, locally maintains some statistics. The cost estimate $C(p)$ of a partition $p$ is as follows:
\begin{equation}
C(p) = N(p) \times Q(p) \times Prob(p)\label{eq:costModel}
\end{equation}
$N(p)$ is the number of points in Partition $p$, $Q(p)$ is the number of queries that overlap $p$, and $Prob(p)$ is the probability that new data and queries land in $p$. $Prob(p)$ depends on the arrival of data and queries during the last round of repartitioning. Note that the workload history is captured via $N$ and $Q$ while $Prob$ is a weighting factor to the cost of this history. The effect of old data can fade with time as discussed in Section~\ref{sec:statisticsMaintaining}.
$Prob(p)$ is estimated as follow:
\begin{equation}
Prob(p) = \frac{R(p)}{R(S)}\label{eq:prob}
\end{equation}
where $R(p)$ and $R(S)$ are the number of data points and queries received by $p$ and all of $S$, respectively, during the last round of repartitioning. $R(S)$ is computed as follows.
\begin{gather}
R(m) = \textstyle\sum_{i=1}^{n_m} R(p_i)\\
R(S) = \textstyle\sum_{i=1}^{|M|} R(m_i)\label{eq:R(S)}
\end{gather}
By substituting Eqn.~\ref{eq:prob} into Eqn.~\ref{eq:costModel}, then:
\begin{equation}
C(p) = \frac{N(p)  Q(p)  R(p)}{R(S)}\label{eq:c(p)}
\end{equation}
The workload of Machine $m$ is computed by:
\begin{equation}
C(m) = \textstyle\sum_{i=1}^{n_m} C(p_i)
\end{equation}
Using Eqn.~\ref{eq:c(p)}, 
\begin{align}
%C(m) =& \frac{(N(p_1) \times Q(p_1) \times R(p_1))}{R(S)}+ ... \nonumber \\
C(m) =& \frac{N(p_1) Q(p_1) R(p_1)}{R(S)}+ \cdots 
+\frac{N(p_{n_m}) Q(p_{n_m}) R(p_{n_m})}{R(S)}\nonumber \\
C(m) =& \frac{\textstyle\sum_{i=1}^{n_m} \{ N(p_i)  Q(p_i)  R(p_i) \} }{R(S)} =  \frac{Num(C(m))}{R(S)} \label{eq:C(m)}
\end{align}
where $Num(C(m))$ is the numerator of Machine $m$'s cost formula. $Num(C(m))$ can be computed locally. In contrast, computing $R(S)$ requires information from all machines in $S$. $R(S)$ is the same for all machines, and hence is computed once using Eqn.~\ref{eq:R(S)} that requires only one number ($R(m)$) from each executor machine. Thus, having $Num(C(m))$ for all machines is enough to compare and rank the machines by cost.

%\vspace{-0.15in}
\section{SWARM Architecture}\label{sec:SWARM}
%\subsection{Architecture}
%\vspace{-0.13in}
\begin{figure}
\centering
\includegraphics[trim=0.1in 0.2in 0in 0in, clip,width=0.82\textwidth]{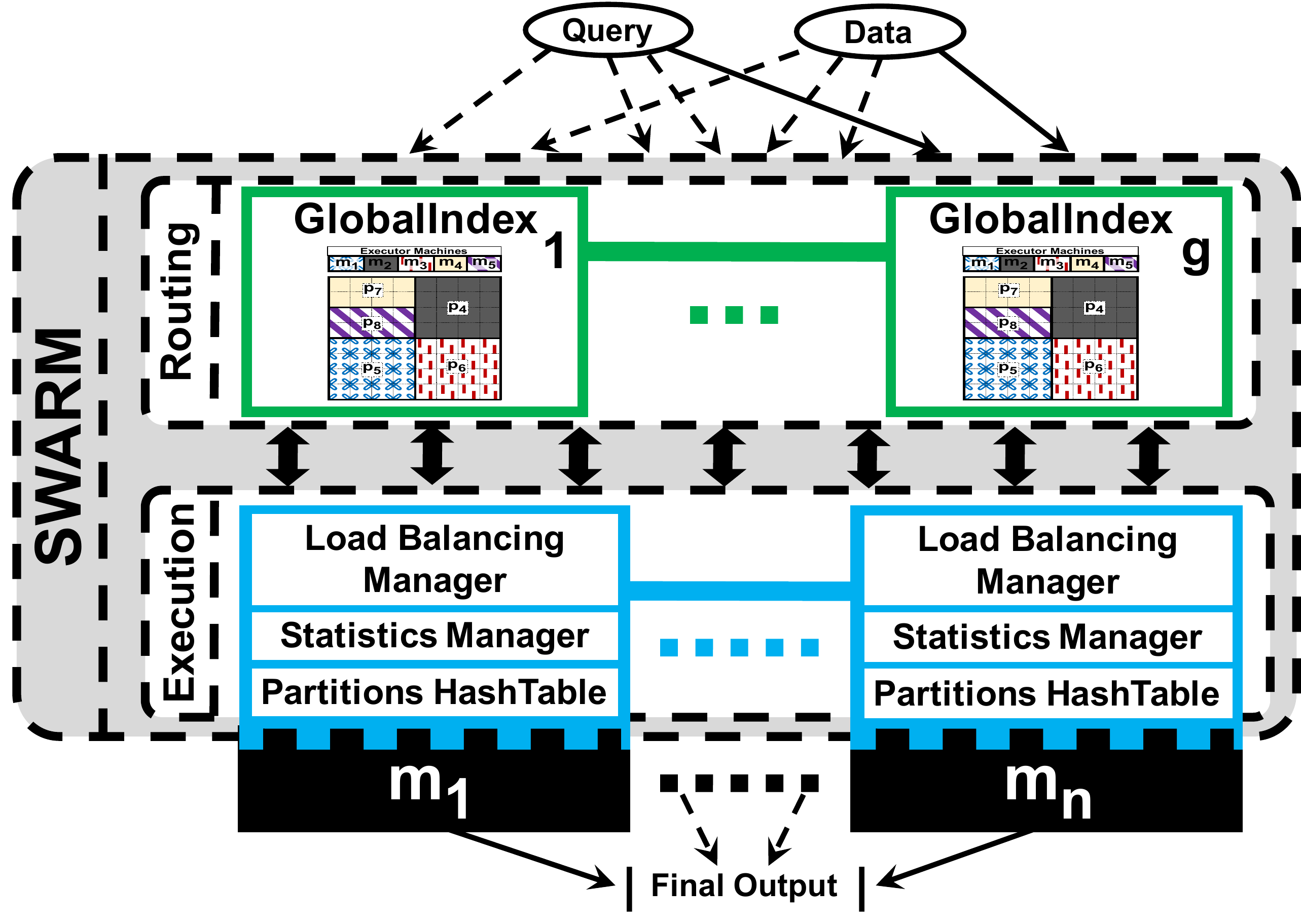}
\caption{\label{fig:architecture} The architecture of SWARM}
\end{figure}
SWARM can work with any distributed streaming system that processes spatial data. 
It does not require changing the original code of the system's executor machines, e.g., their indexes, their way of handling data or processing queries. 
Refer to Figure~\ref{fig:architecture}. SWARM is composed of two layers, the routing layer and the execution load-balancing layer. 
SWARM replaces the partitioning layer of spatial streaming applications (that have a partitioning layer). SWARM is placed on top of the original executor machines $m_1, \cdots, m_n$ and directly receives the incoming streamed data and queries. The routing layer accepts new data points and queries, and routes them to appropriate executor machines. The routing layer has multiple {\em GlobalIndex} machines to avoid bottlenecks. GlobalIndex machines can communicate with each other and with any executor machine. Each GlobalIndex machine has a spatial grid index that divides the whole space into rectangular partitions. Each executor machine is responsible for one or more partitions. Every new data point, say $x$, or query is received by only one GlobalIndex machine that uses the index to identify the partition, say $p_x$, that spatially contains $x$, then routes $x$ to $p_x$'s executor machine. One GlobalIndex machine, termed the Coordinator, has an additional role other than routing. Section~\ref{sec:Indexing} explains that further.
Having GlobalIndex machines reduce the processing overhead, memory usage, and communication among executor machines. The reason is that data points and queries of a partition will be localized in one executor machine. A query will not be sent to all executor machines. Also, communication among executor machines to aggregate the results is  reduced. SWARM uses GlobalIndex machines to adaptively load-balance the workload among executor machines, as in Section~\ref{sec:AdaptiveLoadBalancing}. 

SWARM's second layer contains load-balancing units above the system's original executor machines $m_1, \cdots, m_n$. Each unit has a load-balancing manager, a statistics manager, and a HashTable to index the partitions. Each unit communicates with all other load-balancing and GlobalIndex machines. The unit receives data points and queries only for the partitions that are under its control. The unit's statistic manager of updates its statistics given the new object, as in Section~\ref{sec:statistics}. Moreover, the unit uses a HashTable to identify the partition(s), say $p$, that overlap the received object, as in Section~\ref{sec:Indexing}. Then, the original application code in the executor machine processes this object on $p$. The load-balancing manager computes the cost of its executor machine periodically using the cost model in Section~\ref{sec:costModel} using local statistics. It shares this cost with one GlobalIndex machine. Executor machines with the highest and lowest workloads, say $m_H$ and $m_L$, respectively, are identified. $m_H$ moves part of its workload to $m_L$, as explained in Section~\ref{sec:AdaptiveLoadBalancing}. 

\subsection{Indexing and Initialization}\label{sec:Indexing}
SWARM does not require prior knowledge about the distribution of the incoming data or queries. Initially, SWARM divides the whole spatial area evenly among all executor machines. This section introduces the global and local indexes used by SWARM.

\subsubsection{\textbf{The Global Index}}
\begin{figure}
\centering
\includegraphics[trim=0in 0.1in 0in 0in, clip,width=0.77\textwidth]{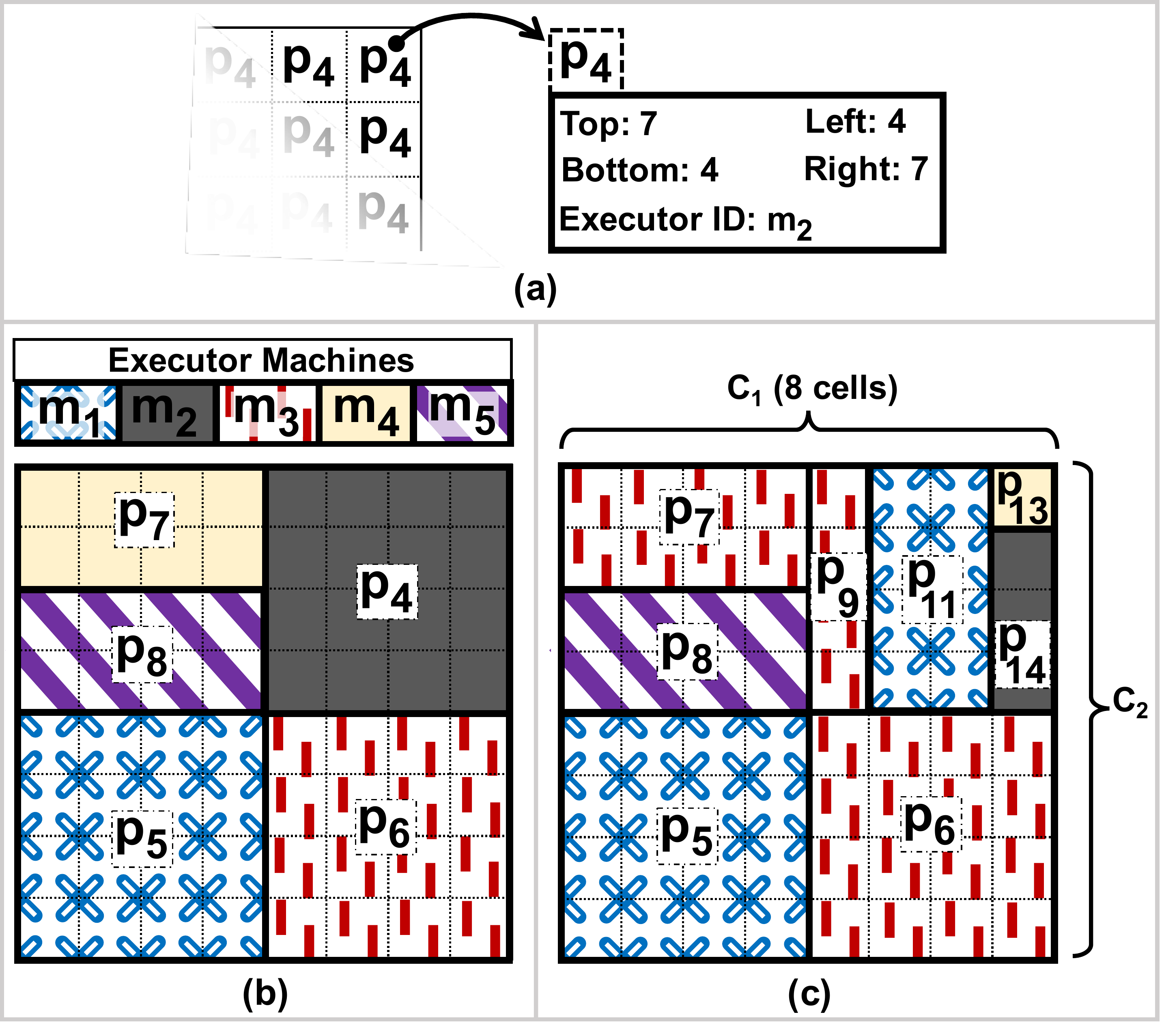}
\caption{\label{fig:globalIndex} SWARM's index for GlobalIndex machines}
\end{figure}

SWARM uses a 2D spatial grid index in each GlobalIndex machine to divide the space into grid cells of a predefined size $C_1 \times C_2$ (refer to Figure~\ref{fig:globalIndex}). This global index replaces the partitioning index of spatial applications that have a partitioning layer. As in Figure~\ref{fig:globalIndex}a, each cell points to a partition that covers this cell. A partition has a unique ID, partition borders, and the ID of the executor machine that handles the partition. Thus, it takes $O(1)$ operations to route an object. Figure~\ref{fig:globalIndex}b gives an example for initial configuration of the index in the GlobalIndex machines of a system with 5 executor machines $m_1, \cdots, m_5$ and 5 partitions $p_1, \cdots, p_5$. The patterns (colours) of the partitions link them to the executor machines that control them.

This index routes the received queries and data points to the responsible executor machine(s). However, one of the GlobalIndex machines, termed the Coordinator, has higher responsibilities for load balancing alongside routing incoming objects. Initially, the whole space is contained in one partition. The Coordinator creates the initial index by recursively splitting the partition with the largest area into two equal sub-partitions, until each executor machine has one partition. The splitting can be horizontal or vertical depending on which side length is longer. Every time a partition is split, say $p_1$, the sub-partitions are given new unique IDs, say $p_2$ and $p_3$. Splitting of a partition stops if a resulting sub-partition is smaller than a cell. Then, the index is shared with all the GlobalIndex machines. Moreover, the Coordinator sends the information about each partition to the executor machine responsible for that partition. The Coordinator is also responsible for identifying the two machines with highest and lowest workloads in each load-balancing round. Every round, the Coordinator receives from each executor machine only two numbers that help determine the cost of all executor machines according to Eqn.~\ref{eq:C(m)}, as explained in  Section~\ref{sec:balancingProcess}. If the Coordinator fails, another GlobalIndex machine takes over as the new Coordinator. This prevents a single point of failure in the system. The decision of choosing a GlobalIndex machine to be the new Coordinator is made by using the Byzantine agreement protocol among all executor machines.

Figure~\ref{fig:globalIndex}c gives a possible configuration of the index when a hotspot appears in the top-right corner of the space. Each load-balancing round, the Coordinator requests from the machine with the highest cost to move some of its partitions' responsibilities to the machine with the lowest cost. As in Figure~\ref{fig:globalIndex}c, the hotspot leads to splitting some partitions and moving others to different machines. An executor machine can be responsible for any number of partitions, e.g., $m_3$ handles 3 partitions. $p_{13}$ has a hotspot, but cannot be split because its size equals a cell's size. Thus, SWARM has an executor machine ($m_4$) responsible for only $p_{13}$. As the hotspot migrates, $m_4$ might become responsible for other partitions. 

Routing a data point is fast as it overlaps only one cell. However, a range query can overlap multiple cells. A naive algorithm for finding which partitions overlap a query can be visiting all cells that overlap this query. Algorithm \ref{alg:queryOverlap} efficiently determines which partitions overlap a query. The algorithm uses SWARM's index and partition structure. 

%\vspace{-0.09in}
\begin{algorithm}
\DontPrintSemicolon
\caption{queryOverlap(Query $q$)}\label{alg:queryOverlap}
Stack<CellCoordinate> \textit{checkCell}\;
List<Partition> \textit{result}\;
indexQuery $iq$ = mapQueryToIndex(\textit{q})\;
\textit{checkCell}.push(CellCoordinate($iq$.left, $iq$.top))\;
\While {!\textit{checkCell}.isEmpty}{
    CellCoordinate $c$ = checkCell.pop\;
    Partition $p$ = gridIndex[$c$.x][$c$.y]\;
    \If{($p$ not in \textit{result}) \&\& ($c$ overlaps $iq$)}{ 
		 \textit{checkCell}.push(CellCoordinate($p$.right+1, $c$.y))\;
		 \textit{checkCell}.push(CellCoordinate($c$.x, $p$.bottom-1))\;
		 \textit{result}.add($p$)\;	
    }
}
\Return{\textit{result}}\;
\end{algorithm}

%\vspace{-0.14in}
Algorithm~\ref{alg:queryOverlap} can skip cells by using the partitions' borders. It adds the coordinates of the cell that overlaps the query's top-left corner to the $checkCell$ stack. If the cell's coordinates ($c$) taken from $checkCell$ overlaps the query, the cell's partition ($p$) is added to the $result$. Also, two cells are added to $checkCell$: 1)~the one after the right border of $p$ on the same row as $c$, and 2)~the one below the bottom border of $p$ and on the same column as $c$. The algorithm recursively takes and adds cells to $checkCell$ until the stack is empty.

\subsubsection{\textbf{The Local Index}}
SWARM adds a local index in each executor machine. Notice that this local index is separate from any other index used in the user's application code. Also, SWARM does not interfere with the application logic in any other way. For example, a user's application for evaluating spatio-textual queries might be using a grid index for maintaining data points or an R-Tree index for storing continuous queries. In other words, user applications can be used as is in executor machines while still using SWARM for load balancing. SWARM's new local index in executor machines receives data points and queries from the GlobalIndex machines. Then, it identifies the required partition for processing the received data points or queries, and forwards them to the user's application for evaluation. 

SWARM adds a HashMap index to each executor machine to easily find the required partition when processing newly received data or queries. Each executor machine uses its HashMap to index the partitions that are under its control. The HashMap key is PartitionID that is the unique identifier of each partition. 

\begin{figure}
\centering
\includegraphics[trim=0.2in 0in 0.12in 0.05in, clip, width=0.75\textwidth]{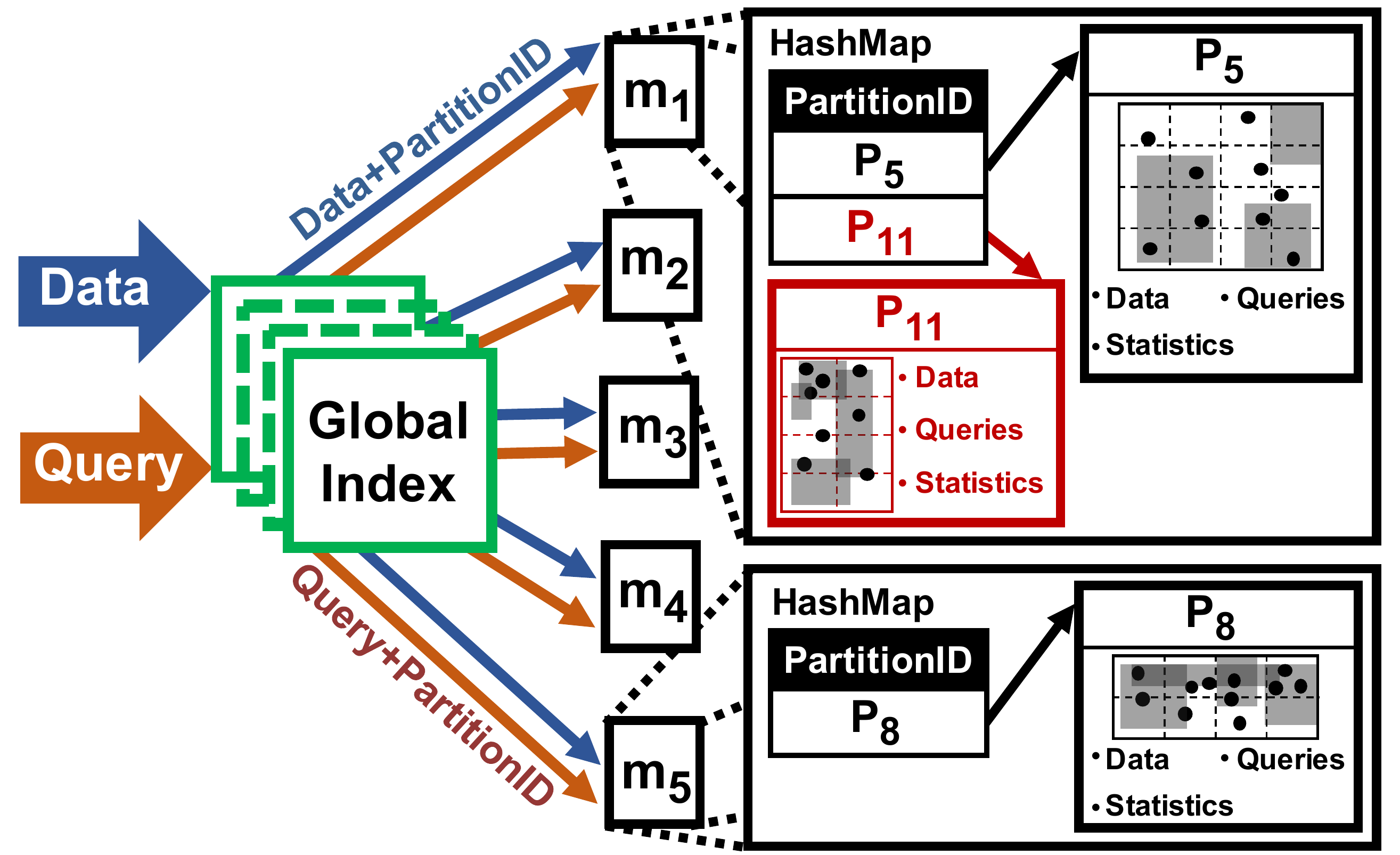}
\caption{\label{fig:machinesPartations} SWARM with the HashMap of $m_1$ and $m_5$}
\end{figure}

Figure~\ref{fig:machinesPartations} illustrates the HashMap of the executor machines $m_1$ and $m_5$. The GlobalIndex machines attach to every received data point or query the PartitionID of the partition that overlaps it. This received data point or query is routed with the PartitionID to the executor machine that is responsible for that partition. The HashMap of $m_5$ contains  partition $p_8$ while $m_1$ has partitions $p_5$ and $p_{11}$. 
%Each time an executor machine receives a data point or query from a GlobalIndex machine, it directly uses the attached PartitionID with the HashMap to find the structure of corresponding partition in $O(1)$. 
The structure of a partition maintains its metadata (PartitionID, position in space, and size), data points, queries, and statistics. The partition's data points and queries are maintained using the code of the original application.

\subsection{Collecting and Maintaining Statistics}\label{sec:statistics}
Collecting statistics in distributed streaming systems is challenging because the data arrives in high continuous volumes. Moreover, most applications need real-time processing for each data point with minimum latency. Thus, a feasible technique for collecting and maintaining Statistics in distributed streaming systems should require minimum number of updates. Also, each partition should maintain its statistics locally without the need to communicate with other machines. SWARM achieves this by maintaining minimum local statistics that are enough to make all load balancing decisions locally. SWARM maintains the statistics in a simple multidimensional array in memory, and takes advantage of cache prefetching as all the needed statistics to perform one cost calculation are located next to each other in memory. 
%The remaining of this section discusses in details the process of collecting and maintaining statistics in SWARM.

\subsubsection{\textbf{SWARM's Statistics}}\label{sec:SWARMStatistics}
SWARM maintains the minimum statistics needed for using the cost model. Space is divided into a grid of small cells that are aligned with the grid index of GlobalIndex machines. The arrangement of cells that cover a partition are passed to the responsible executor machine with the partition's metadata. We use Partition $p_{11}$ in Figures~\ref{fig:globalIndex}c and~\ref{fig:machinesPartations} to illustrate how SWARM  maintains the statistics. 

\begin{figure}
\centering
\includegraphics[trim=0.1in 0in 0in 0in, clip,width=0.7\textwidth]{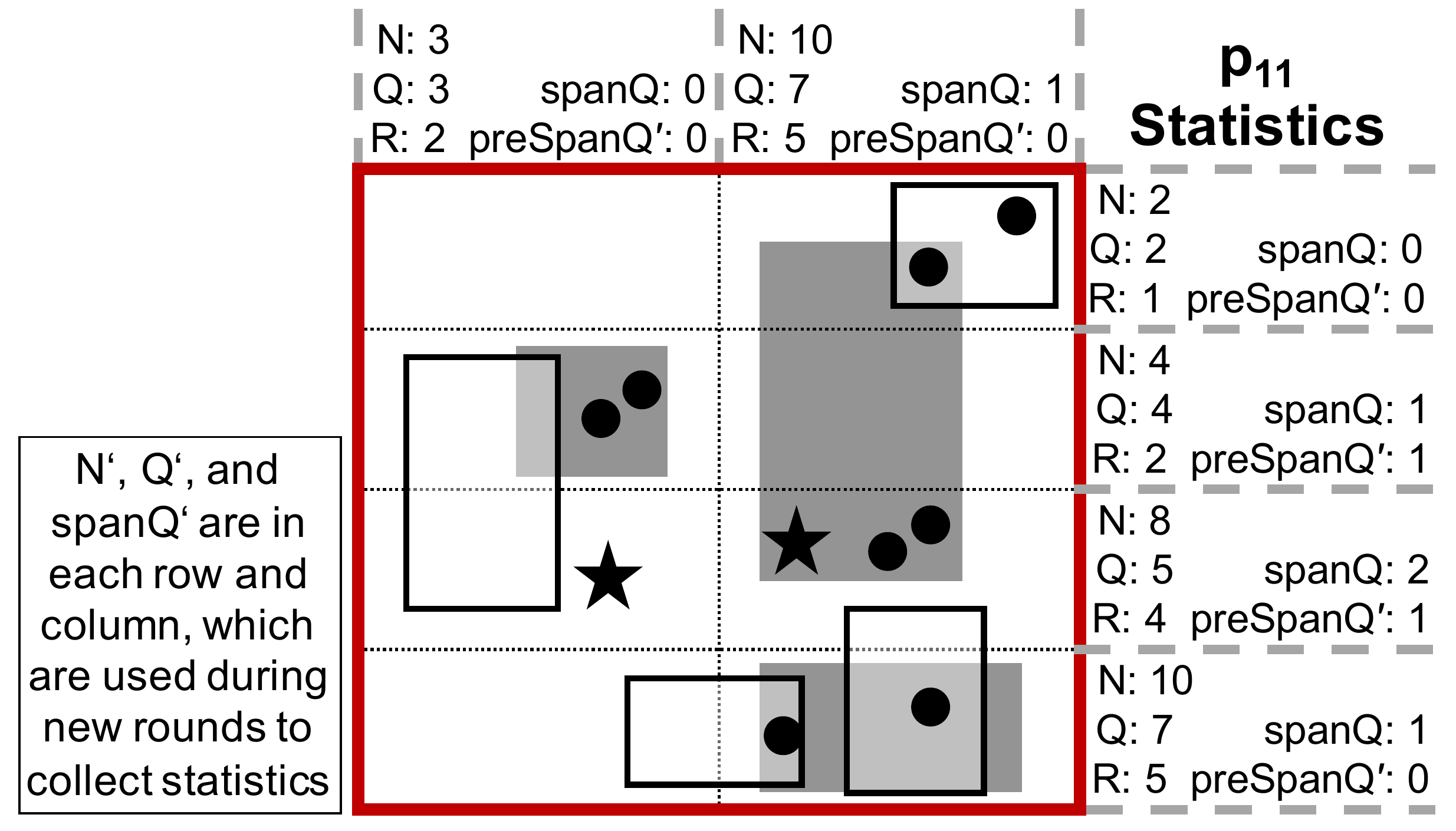}
\caption{\label{fig:statistics} SWARM statistics for partition $p_{11}$}
\end{figure}

Figure~\ref{fig:statistics} gives the maintained statistics in  $p_{11}$ at the end of a load-balancing round. The dots and rectangles represent the positions of the data points and the query ranges in $p_{11}$, respectively. The stars and the gray rectangles mark the data points and the queries received in the last round of load balancing, respectively. $p_{11}$ has a 4X2 cell matrix. 
%The indexes of the rows start form top to bottom, while it is from left to right for the columns. 
SWARM maintains in each row and column 5 statistics, 3 of which are cumulative. Row~$i$'s (Column~$j$'s) cumulative statistics represent the total from the uppermost row (leftmost column) until Row~$i$ (Column~$j$), respectively. The 5 maintained statistics in each row and column are: (1)~$N$: the cumulative number of data points, (2)~$Q$: the cumulative number of queries, (3)~$R$: the cumulative number of data points and queries received during the last round of load balancing, (4)~$spanQ$: the number of queries whose ranges span from the previous row/column, and (5)~$preSpanQ'$: the number of queries received during the last load balancing round whose ranges span from the previous row/column. To illustrate, refer to Row~3 of $p_{11}$ in Figure~\ref{fig:statistics}. All cumulative statistics reflect the objects in the first three rows. There are 8 data points ($N$) and 5 queries ($Q$). Two data points and two queries are received during the last round, hence $R=4$. Two queries span from the second row ($spanQ=2$). However, only one of them is received during last round ($preSpanQ'=1$). 

SWARM uses these statistics for load-balancing purposes. The overall statistics of a partition $p$ ($N(p)$, $Q(p)$, and $R(p)$) are the ones in the last row/column. The statistics are only updated at the end of a load-balancing round to avoid the overhead of updating almost all the statistics whenever a new data point or query arrives.  Additional three statistics, termed {\em Statistics Collectors}, for each row and column are introduced, namely $N'$, $Q'$, and $spanQ'$. {\em Statistics Collectors} are used to update the statistics at the end of a round. They reduce the number of updates per received data point or query. The next section presents how these \textit{Statistic Collectors} are updated and used for maintaining the statistics.

\subsubsection{\textbf{Maintaining the Statistics}}\label{sec:statisticsMaintaining}
SWARM needs to have a small number of updates when receiving a data point or query. When a new data point arrives, SWARM updates only two of a partition's \textit{Statistics Collectors}. However, when a new query arrives, SWARM updates the \textit{Statistics Collectors} of the rows and columns that overlap the query. Having more statistics to update will not affect the performance because the arrival rate of data is much higher than that of queries.

Three \textit{Statistics Collectors}, $N'$, $Q'$, and $spanQ'$, are used in each row/column to count different types of received objects during the most recent round of load balancing. $N'$ and $Q'$ count the new data points and queries, respectively. $spanQ'$ counts the number of queries that their ranges span from the previous row/column. When a new data point arrives, SWARM increments $N'$ of the row and the column containing the data point. When a new query arrives, SWARM increments both $Q'$ of the row and the column that overlap the top-left corner of the query, and $spanQ'$ of the rows and the columns that overlap the query excluding the row and the column overlapping the top-left corner of the query. 

To conclude a load-balancing round, SWARM uses the {\em Statistics Collectors} to update all remaining statistics as follows. Let $i \geq 0$ be a row/column index. Then, the statistics are updated as follows:
%\vspace{-0.2in}
\begin{align*}
N(i) = N(i) + \textstyle\sum_{j=0}^i N'(j)&~~~,~~~~
Q(i) = Q(i) + \textstyle\sum_{j=0}^i Q'(j)\\
R(i) = \textstyle\sum_{j=0}^i N'(j) &+ \textstyle\sum_{j=0}^i Q'(j)\\
spanQ(i) = spanQ(i) + spanQ'(i)&~~~,~~~~
preSpanQ'(i) = spanQ'(i)
\end{align*}

%\vspace{-0.10in}
\begin{algorithm}[H]
\DontPrintSemicolon
\caption{updateStat(\textit{PartitionID}, \textit{rowOrColumn})}\label{alg:statistics}
\textit{stat}$[~][~]$ = partitionsHashMap.get(\textit{PartitionID})\;
\pushline\nonl.statistics(\textit{rowOrColumn}) \Comment{\textnormal{Multidimensional array}}\;
\popline int $sumN' = 0$\;
int $sumQ' = 0$\;
\For {$i = 0$ \textnormal{to Num of \textit{rowOrColumn} in \textit{PartitionID}}}{
    $sumN'~{+}{=}~$\textit{stat}$[N'][i]$\;
	$sumQ'~{+}{=}~$\textit{stat}$[Q'][i]$\;
	\textit{stat}$[N'][i] = 0$ \Comment{\textnormal{Reset current $N'$}}\;
	\textit{stat}$[Q'][i] = 0$ \Comment{\textnormal{Reset current $Q'$}}\;
    \textit{stat}$[N][i]~{+}{=}~sumN'$\;
    \textit{stat}$[Q][i]~{+}{=}~sumQ'$\;
    \textit{stat}$[$\textit{preSpanQ}$'][i] =$\textit{stat}$[spanQ'][i]$\;
    \textit{stat}$[spanQ][i]~{+}{=}~$\textit{stat}$[spanQ'][i]$\;
    \textit{stat}$[spanQ'][i] = 0$ \Comment{\textnormal{Reset current $spanQ'$}}\;
    \textit{stat}$[R][i] = sumN' + sumQ'$\;
}
\end{algorithm}

SWARM uses an efficient algorithm that utilizes the fact that the summations in the equations can be carried out from one row/column to another. There is no need to compute the summations from scratch each time, i.e., $O(n!)$. With only one addition, we produce the statistics of the next row/column from these of the  previous row/column. Algorithm~\ref{alg:statistics} illustrates how to update the statistics of a partition by passing once through the partition's rows and columns, i.e., $O(n)$. This algorithm runs as a separate task in the background. Note that all \textit{Statistics Collectors} are reset to $0$ to be ready for collecting the statistics of the next round of load balancing.

\begin{figure}
\centering
\includegraphics[trim=0.1in 0in 0in 0in, clip,width=\textwidth]{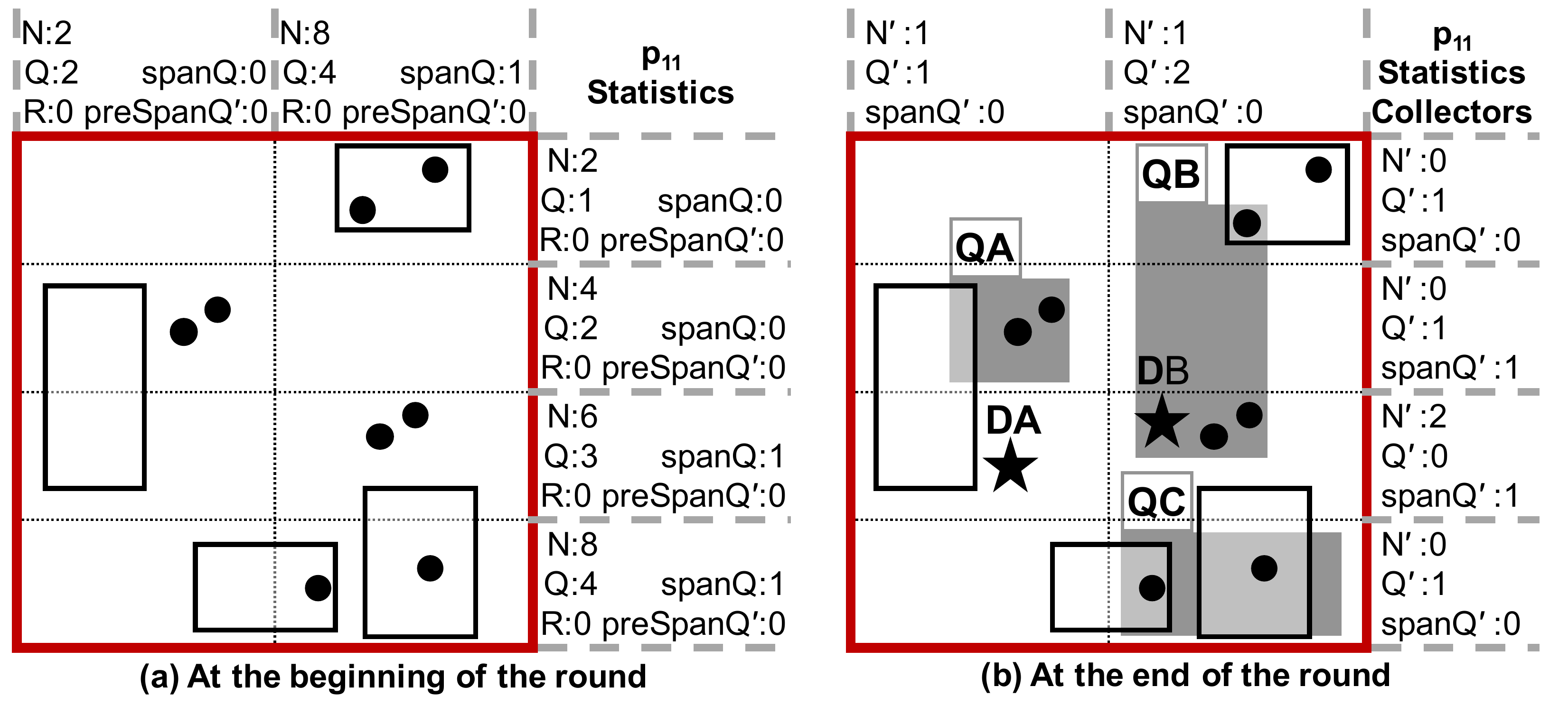}
\caption{\label{fig:statisticsCollecting} Updating partition $p_{11}$'s \textit{Statistics Collectors}}
\end{figure}

%\vspace{-0.13in}
Figure~\ref{fig:statisticsCollecting} illustrates the statistics of Partition $p_{11}$ while receiving new data points and queries. Figure~\ref{fig:statisticsCollecting}a illustrates the positions of the data points and the ranges of the queries in $p_{11}$ at the beginning of a new load balancing round. Also, it shows the current state of the maintained statistics as discussed in Section~\ref{sec:SWARMStatistics}. The \textit{Statistics Collectors} are all set to $0$ at the beginning of the round. Figure~\ref{fig:statisticsCollecting}b shows the \textit{Statistics Collectors} at the end of the round after receiving 2 new data points and 3 new queries. During the load-balancing round, the two data points $D_A$ and $D_B$ are received first. Both data points are in the third row ($Row_2$), but one of them is in the first column ($Col_0$) while the other is in the second column ($Col_1$). $N'(Row_2)$ is incremented twice while $N'(Col_0)$ and $N'(Col_1)$ are each incremented once. Then, Queries $Q_A$, $Q_B$, and $Q_C$ arrive into $p_{11}$ in this order. The upper-left corner of $Q_A$ is in the cell that overlaps $Col_0$ and $Row_1$. Also, the range of $Q_A$ is contained within one cell. Thus, only $Q'(Row_1)$ and $Q'(Col_0)$ are incremented. $Q_B$ starts in $Row_0$ and spans through $Row_1$ and $Row_2$. Thus, $spanQ'(Row_1)$ and $spanQ'(Row_2)$ are incremented in addition to the increment of $Q'(Row_0)$ and $Q'(Col_1)$. At the end of the round, \textit{Statistics Collectors} are used to update the statistics using Algorithm~\ref{alg:statistics}. The results of the updated statistics are given in Figure~\ref{fig:statistics}.

Notice that the target of SWARM is not to count the actual number of data points but rather to track the change in the spatial data workload. To diminish the effect of old data gradually, the number of data points $N$ is divided by $2$ before it is updated in each round of load balancing. This is to reduce the effect of old data points on the current spatial distribution. In distributed streaming systems that support historical queries, SWARM needs to be informed about data expiration to update $N$ accordingly. 

\subsubsection{\textbf{Correctness of the Statistics}}
In this section, we prove the correctness of the statistics that SWARM collects and maintains about data points and queries. To show that, we need to prove that the maintained statistics always represent the true number of data points and queries without any over- or under-counting. 

\begin{figure}[h]
\centering
\includegraphics[width=0.45\textwidth]{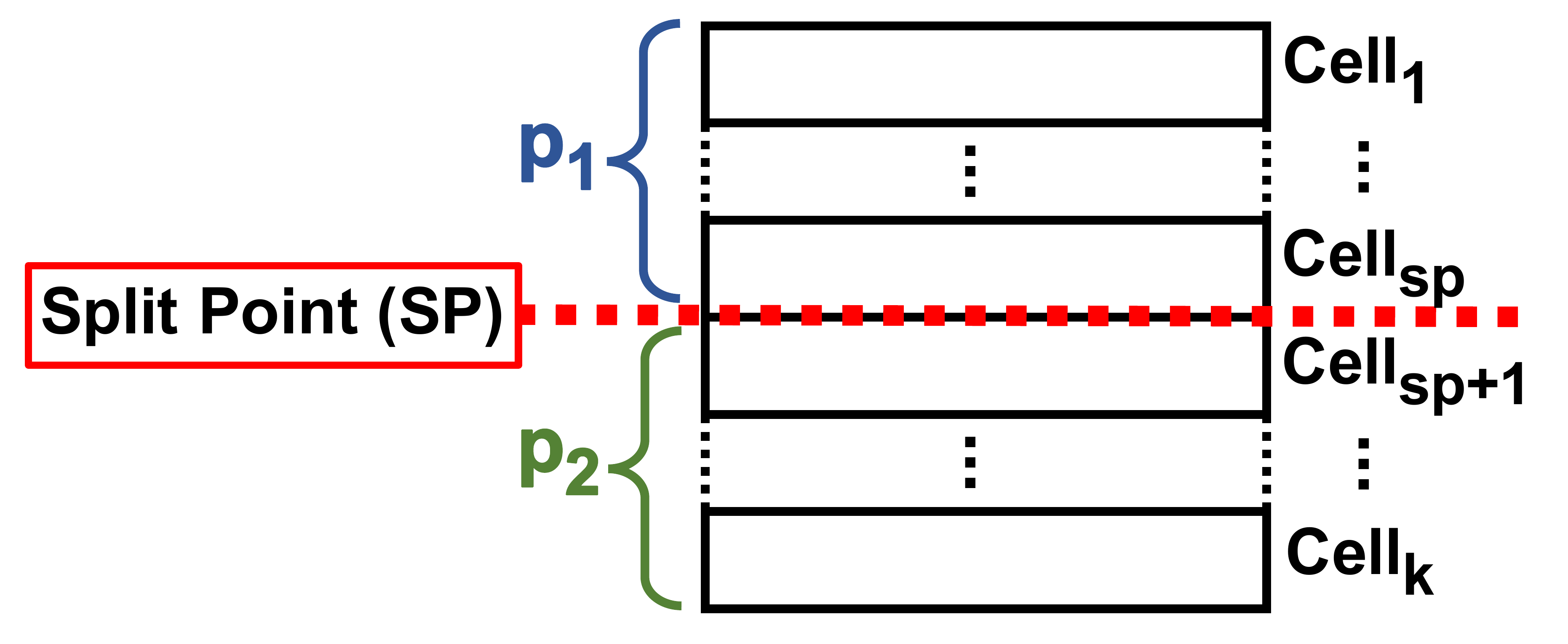}
\caption{\label{fig:proofStat} Partition with $k$ cells (rows)}
\end{figure}

First, we prove the correctness of the statistics for data points. Assume that we have a partition that has $k$ rows and only one column. This results in $k$ cells in total as in Figure~\ref{fig:proofStat}.

Let $i$ be the row number of a cell, where $1 \leq i \leq k$, $n(i)$ be the true number of data points in $cell_i$, and $N(i)$ be the cumulative number of data points that SWARM maintains in $row_i$. $n(i)$ can be obtained by simply counting the number of data points within $cell_i$. As mentioned before, the cumulative number $N(i)$ is computed from top to down for horizontal divisions. Therefore, $N(i)$ can be computed as follows:
\begin{equation*}
N(i) = \textstyle\sum_{j=1}^i n(j)
\end{equation*}
In the initial case where $k=1$, there is only one cell with $n(1)$ data points, hence $N(1) = n(1)$. For $k=2$, $N(2) = n(1) + n(2)$. We can derive the number of data points in $cell_2$ as $n(2) = N(2) - N(1)$. In general, let us assume a partition as illustrated in Figure~\ref{fig:proofStat} that has a split point $sp$, where $1<=sp<=k$, that divides the partition into two partitions, say $p_1$ and $p_2$. $n(p_i)$ represents the true number of data points in Partition $p_i$. Therefore, the number of data points in each partition can be computed as follows:
\begin{align*}
n(p_1) &= n(1)+n(2)+...+n(sp)\\
\therefore n(p_1) &=\textstyle\sum_{j=1}^{sp} n(j) = N(sp)\\
n(p_2) &= n(sp+1)+n(sp+2)+...+n(k) = \textstyle\sum_{j=sp+1}^{k} n(j)\\
\therefore n(p_2) &= \textstyle\sum_{j=1}^{k} n(j) - \textstyle\sum_{z=1}^{sp} n(z) = N(k) - N(sp)
\end{align*}
Therefore, this shows that the computed statistics, i.e., $N$, is equal to the true number of data points, i.e., $n$. An analogous proof can be used to show that $N$ is also correct when dividing cells vertically, and $N$ is cumulatively computed from left to right. 

Now, we prove the correctness of the maintained statistics about queries by using the same setup of  Figure~\ref{fig:proofStat}. Given the input query boundaries, we can extract the exact count of  all queries in each grid $cell_i$ by maintaining four variables, namely, $q_s, q_e, q_{se}$ and $q_o$ ($s$ stands for start, $e$ stands for end, $se$ stands for start and end, and $o$ stands for overlap, as explained below). Let $q_s(i)$ be the number of queries whose upper boundary intersects $cell_i$ and whose lower boundary intersects another cell. Furthermore, let $q_e(i)$ be the number of queries whose lower boundary intersects $cell_i$ and whose upper boundary intersects another cell. Let $q_{se}(i)$ be the number of queries whose upper and lower boundaries intersect $cell_i$. Finally, let $q_o(i)$ be the number of queries whose upper and lower boundaries do not intersect $cell_i$ but their ranges overlap $cell_i$. Therefore, the true number, $q(i)$, of queries that intersect $cell_i$ is the sum of these four variables, i.e.,
\begin{equation}
q(i) = q_s(i) + q_e(i) + q_{se}(i) + q_o(i)
\label{eq:q_true}
\end{equation}
Now, we extend the formula above to compute the true number of queries that overlap a sub-partition, i.e., one column of cells that starts from Row $u$ and ends in Row $l > u$. Let that number of $q(u,l)$. We need to avoid double counting of a query that overlaps multiple cells. $q(u,l)$ is equal to the true number of queries in Row $u$ and only queries that start in any row from row $u+1$ up to row $l$, no matter where these queries end. For rows after $u$, only counting queries that start in any cell will exclude recounting any query that span over multiple cells. Therefore, $q(u,l)$ can be computed using the four variables as follow:
\begin{equation}
%q(u,l) = q(l) + \textstyle\sum_{j=u}^{l-1} q_e(j) + q_{se}(j)
q(u,l) = q(u) + \textstyle\sum_{j=u+1}^{l} (q_s(j) + q_{se}(j))\label{eq:q_u_i_true}
\end{equation}

We need to demonstrate that the statistics gathered by SWARM when counting the number of queries is equal to the true number, i.e., $q(u,l)$. Refer to Figure~\ref{fig:proofStat} for illustration. As in the figure, we have a partition that has $k$ cells starting from Cell 1 at the top down to Cell $k$ at the bottom. To maintain query statistics, for each Row $i$, where $1 \leq i \leq k$, of a partition, SWARM maintains only two statistics per row,  namely, $Q(i)$ and $Q_{span}(i)$. $Q(i)$ is the cumulative number of queries from the first row of the partition, i.e., Row $1$, to Row $i$. Therefore, $Q(i)$ directly represents the number of queries that start at any row from the beginning of the partition until Row $i$. Recounting of queries can happen by considering queries that only end or overlap any of the cells as they are already counted where they started. Thus, they are excluded from $Q(i)$ as follows:
\begin{equation}
Q(i) = \textstyle\sum_{j=1}^{i} (q_s(j) + q_{se}(j))\label{eq:Q}
\end{equation}

Let $Q_{span}(i)$ be the number of queries that extend (span) from an upper row, say Row $(i-1)$, to Row $i$. Thus, $Q_{span}(i)$ represents the number of queries that overlap or start without ending in Row $(i-1)$. Note that $Q_{span}(1)$ will always equal $0$ because there are no queries that extend from outside the partition to the first row. Thus, $Q_{span}(i)$ can be formulated from the true numbers as follows:
\begin{align*}
Q_{span}(i) &= 
		\begin{cases}
			0,~~~~~~~~~~~~~~~~~~~~~~~~~~\textnormal{if }i = 1\\
			q_s(i-1) + q_o(i-1),~\textnormal{otherwise}
		\end{cases}
\end{align*}
Although $Q_{span}(i)$ depends on the variables of the previous row $(i-1)$ in SWARM, there is another equivalent way of computing $Q_{span}(i)$ with the variables from Row $i$ that makes the proof easier to follow. Note that any query that overlap Row $(i-1)$ or starts without ending in Row $(i-1)$ definitely extends to Row $i$ and this query's range either ends at Row $i$ or overlaps Row $i$ and continues to the next row below Row $i$. This can be reflected in the formula for calculating $Q_{span}(i)$ as follows:
\begin{align}
\because q_s(i-1) + q_o(i-1) &= q_e(i) + q_o(i) \nonumber\\
\therefore Q_{span}(i) &= q_e(i) + q_o(i) 
\label{eq:Qspan}
\end{align}
This equivalent equation for calculating $Q_{span}(i)$ is correct also in the case when $i=1$ because both $q_e(1)$ and $q_o(1)$ are always equal $0$.

In the initial case, i.e., when $k=1$, and there is only one cell in the partition with $q(1)$ queries, $Q(1) = q_s(1) + q_{se}(1) = q(1)$ and $Q_{span}(1) = 0$. This is correct because $q_s(1)$, $q_e(1)$ and $q_o(1)$ are all equal $0$ as $cell_1$ covers the whole partition and every query definitely starts and ends in this cell. For $k=2$, using Eqns.~\ref{eq:Q} and \ref{eq:Qspan}, $Q$ and $Q_{span}$ for Rows 1 and 2 are computed as follows:
\begin{align*}
Q(1) &= q_s(1) + q_{se}(1)\\
Q(2) &= Q(1) + q_s(2) + q_{se}(2)\\
Q_{span}(1) &= q_e(1) + q_o(1) = 0\\
Q_{span}(2) &= q_e(2) + q_o(2)
\end{align*}
Notice that when $k=2$, there are only the following three possible sub-partitions: a partition that has $cell_1$ only, $cell_2$ only, or $cell_1$ and $cell_2$. The computation of the true numbers can be computed using Eqn.~\ref{eq:q_u_i_true} as follows:
\begin{align*}
q(1,1) &= q(1) = q_s(1) + q_e(1) + q_{se}(1) + q_o(1)\\
\therefore &= q_s(1) + 0 + q_{se}(1) + 0 = Q(1)\\
q(1,2) &= q(1) + q_s(2) + q_{se}(2)\\ 
&= q_s(1) + q_e(1) + q_{se}(1) + q_o(1) + q_s(2) + q_{se}(2)\\
&= q_s(1) + 0 + q_{se}(1) + 0 + q_s(2) + q_{se}(2)\\ 
\therefore &= Q(1) + q_s(2) + q_{se}(2) = Q(2)\\
q(2,2) &= q(2) = q_s(2) + q_e(2) + q_{se}(2) + q_o(2)\\
\therefore &= Q(2) - Q(1) + Q_{span}(2)
\end{align*}
Note that the maintained statistics are enough to compute the true number of queries in all possible sub-partitions when $k=2$.
% For $k=3$, the computation is as follow:
% \begin{align*}
% Q(1) &= q_s(1) + q_{se}(1)\\
% Q(2) &= Q(1) + q_s(2) + q_{se}(2)\\
% Q(3) &= Q(2) + q_s(3) + q_{se}(3)\\
% Q_{span}(1) &= q_e(1) + q_o(1) = 0\\
% Q_{span}(2) &= q_e(2) + q_o(2)\\
% Q_{span}(3) &= q_e(3) + q_o(3)\\
% q(1,2) &= q(1) + q_s(2) + q_{se}(2)\\ 
% &= q_s(1) + q_e(1) + q_{se}(1) + q_o(1) + q_s(2) + q_{se}(2)\\
% &= q_s(1) + 0 + q_{se}(1) + 0 + q_s(2) + q_{se}(2) = Q(2)\\
% q(2,3) &= q(2) + q_s(3) + q_{se}(3)\\
% &= q_s(2) + q_e(2) + q_{se}(2) + q_o(2) + q_s(3) + q_{se}(3)\\
% &= Q(3) - Q(1) + Q_{span}(2) \\
% \end{align*}
Refer to Figure~\ref{fig:proofStat} for illustration.
For cases $k>2$, assume that a partition has a split point $sp$, where $1<=sp<=k$, that divides the partition into two sub-partitions, say $p_1$ and $p_2$. SWARM's query statistics can be computed using the Eqns.~\ref{eq:Q} and \ref{eq:Qspan} as follows:
\begin{align*}
Q(sp) &= \textstyle\sum_{j=1}^{sp} (q_s(j) + q_{se}(j))\\
Q(k) &= \textstyle\sum_{j=1}^{k} (q_s(j) + q_{se}(j))\\
Q_{span}(sp+1) &= q_o(sp+1) + q_e(sp+1)
\end{align*}
The true number of queries in the partition $p_1$ is computed using the Eqns.~\ref{eq:q_u_i_true} and \ref{eq:q_true} as follow:
\begin{align*}
q(p_1) &= q(1,sp) = q(1) + \textstyle\sum_{j=2}^{sp} (q_s(j) + q_{se}(j))\\ 
&= q_o(1) + q_e(1) + q_{se}(1) + q_s(1) + \textstyle\sum_{j=2}^{sp} (q_s(j) + q_{se}(j))\\
&= q_o(1) + q_e(1) + \textstyle\sum_{j=1}^{sp} (q_s(j) + q_{se}(j))\\
\therefore &= 0 + 0 + \textstyle\sum_{j=1}^{sp} (q_s(j) + q_{se}(j)) = Q(sp)
\end{align*}
Notice that the computed statistic $Q(sp)$ is exactly equal the true number of queries in $p1$, i.e., $q(p_1)$. The true number of queries in the Partition $p_2$ is computed as follows:
\begin{align*}
q(p_2) &= q(sp+1,k) = q(sp+1) + \textstyle\sum_{j=sp+2}^{k} (q_s(j) + q_{se}(j))\\ 
&= q_o(sp+1) + q_e(sp+1) + q_{se}(sp+1) + q_s(sp+1) \\
	&~~~~~~+ \textstyle\sum_{j=sp+2}^{k} (q_s(j) + q_{se}(j))\\
&= q_o(sp+1) + q_e(sp+1) + \textstyle\sum_{j=sp+1}^{k} (q_s(j) + q_{se}(j))\\
\therefore q(p_2) &= Q_{span}(sp+1) + \textstyle\sum_{j=sp+1}^{k} (q_s(j) + q_{se}(j))\\
&= Q_{span}(sp+1) + \textstyle\sum_{j=1}^{k} (q_s(j) + q_{se}(j))\\
	&~~~~~~- \textstyle\sum_{j=1}^{sp} (q_s(j) + q_{se}(j))\\
\therefore q(p_2) &= Q_{span}(sp+1) + Q(k) - Q(sp)
\end{align*}
Therefore, SWARM's statistics ($Q$ and $Q_{span}$) are necessary and sufficient to compute the true number of queries. The same proof can be used to show that SWARM's statistics are correct by using the computed cumulative number $Q$ from left to right when dividing partitions vertically.

For SWARM's Statistic $R$, 
notice that $R$ represents the cumulative number of the newly received data points and queries.
Thus, the proof of correctness for $R$ is that same as the ones for the data points and the queries proofs explained above. 
However, in the proofs, $Q_{preSpan}$ is to used instead of $Q_{span}$, where the former represents the span of only the new queries.

\subsection{Adaptive Load Balancing}\label{sec:AdaptiveLoadBalancing}
SWARM adopts a lazy repartitioning mechanism to balance the workload. It does not over-react to transient changes in the workload so as not to overwhelm the system by excessive load balancing activities. It rebalances the workload only if rebalancing will enhance the system's throughput and reduce its execution latency. Each round of load balancing, SWARM considers only the machine, say $m_H$, with the highest cost for workload reduction. Moreover, workload reduction can only happen by moving some of the partitions' responsibilities to the machine, say $m_L$, with the lowest cost. 

\subsubsection{\textbf{Minimizing Communication by Maximizing Local Computations and Local Decision Making}}\label{sec:balancingProcess}
Most applications that use distributed streaming systems heavily depend on the network bandwidth and connection availability between machines. One of the main objectives of SWARM is to minimize communication. This objective is achieved by: 1)~locally collecting and maintaining statistics with minimum overhead as discussed in Section~\ref{sec:statistics}. 2) The cost model is designed to allow mostly local computations. 3) SWARM breaks the process of load balancing into stages of local decision making. The amount of local computations increases with each stage while the number of machines performing the computations decreases significantly. SWARM uses an asynchronous approach for communicating and applying new partitioning plan among executor machines.

\begin{figure}
\centering
\includegraphics[width=0.75\textwidth]{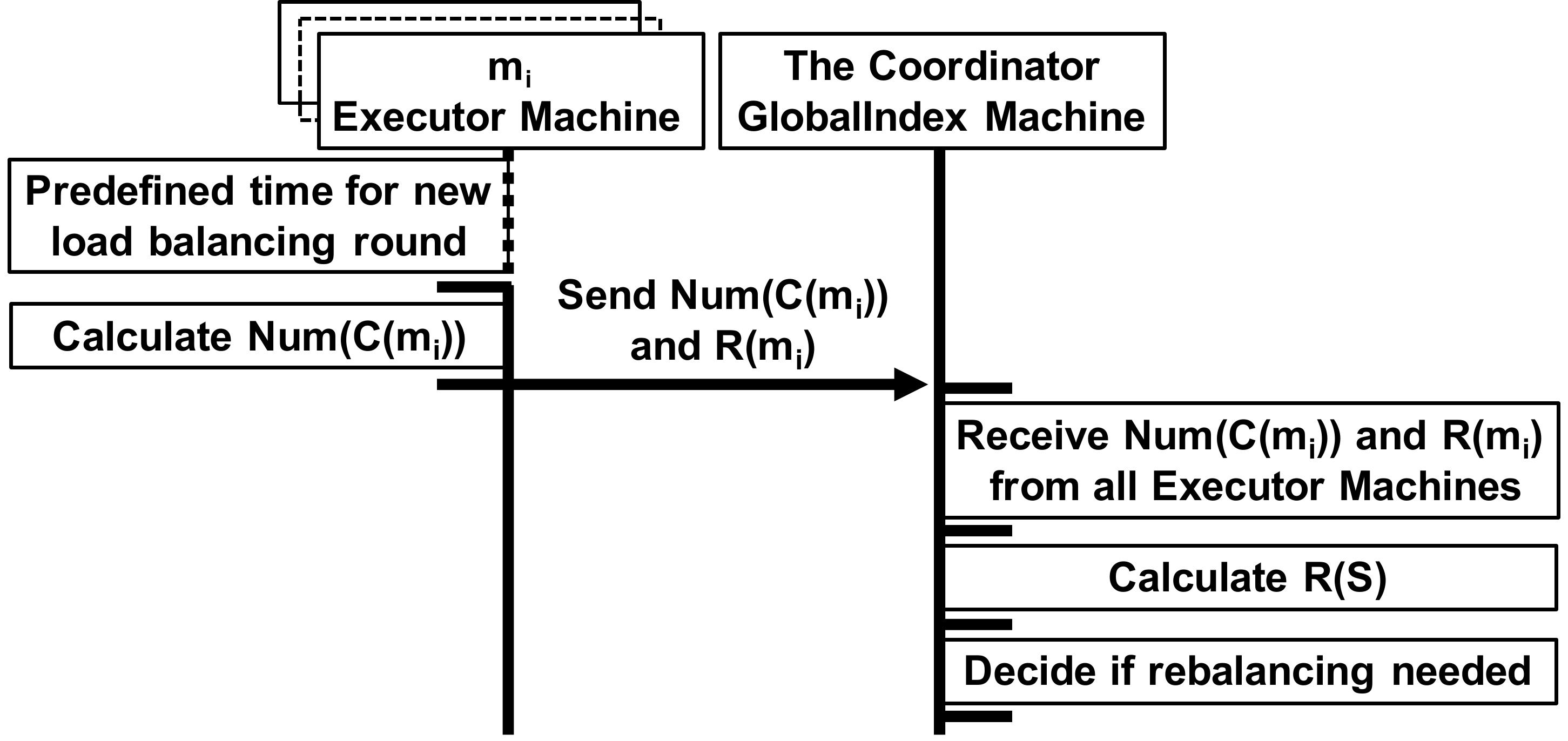}
\caption{\label{fig:balancingProcess1} Workflow of load balancing decision}

\end{figure}

Figure~\ref{fig:balancingProcess1} illustrates communications, local computations, and local decision-making between one of the executor machines, say $m_i$, and the Coordinator. Load balancing is triggered periodically in all executor machines, e.g., every 15 seconds. At the beginning, each executor updates its statistics since the last round of load balancing using Algorithm~\ref{alg:statistics}. Then, the numerator part of the cost equation, i.e., $Num(C(m_i))$, is calculated using local statistics. $Num(C(m_i))$ is sent to the Coordinator along with $R(m_i)$. The Coordinator computes $R(S)$ using Eqn.~\ref{eq:R(S)} after receiving $Num(C(m_i))$ and $R(m_i)$ from all executor machines. Then, the Coordinator decides the best load balancing action.

\begin{figure}
\centering
\includegraphics[width=0.65\textwidth]{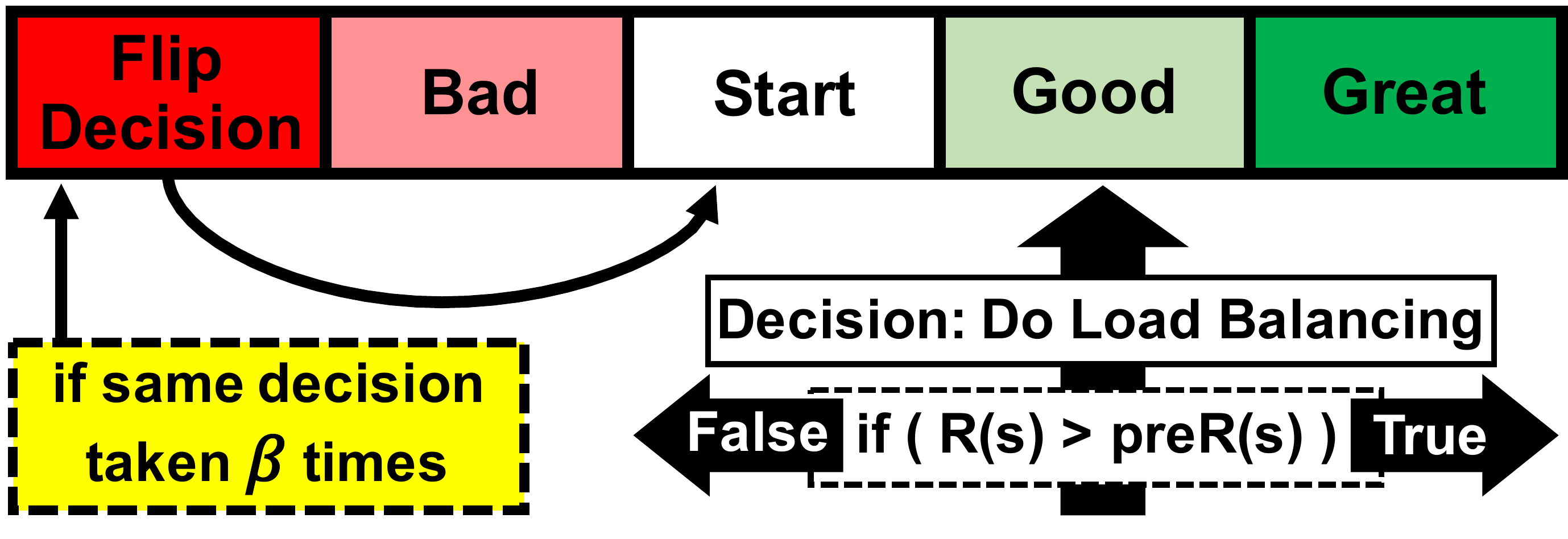}
\caption{\label{fig:decision} Decision mechanism for load balancing}
\end{figure}
Load balancing in SWARM is an iterative process. As in Figure~\ref{fig:decision}, the decision mechanism has five stages, and has a structure that points to the current stage, and that stores the decision made in the previous round. In each round, the Coordinator makes one of two possible load-balancing decisions, either to rebalance the workload, or to simply do nothing. The Coordinator applies the previous load-balancing decision each round unless the \textit{Flip Decision} (leftmost) stage is reached. The leftmost stage flips the decision and resets the stage pointer to the \textit{Start} stage. Initially, the \textit{Start} (middle) stage is selected, and the previous decision is set to "Do Nothing". 

In every load balancing round, the Coordinator moves the pointer to the right if the throughput has enhanced ($R(S) > preR(S)$). Otherwise, it moves the pointer to the left. Thus, moving to the right indicates that the overall throughput is enhancing and the decision performed in the previous round (iteration) has proven correct. This mechanism insures that the current decision continues to be carried over into future rounds until it is ineffective, and in this case, it is flipped. This avoids over-reacting to the system's transient fluctuations in performance.

When the same decision was taken for $\beta$ number of times (e.g., 20), the stage pointer is forced to move to the \textit{Flip Decision} stage. This is to avoid sticking with one decision for a long time while this decision is making the system stay in a sub-optimal partitioning plan, e.g., if the Coordinator decides to do nothing each round because the throughput keeps increasing in one round and decreasing in the next. Forcing the system to try rebalancing may lead to better throughput. Otherwise, the decision will be flipped again to do nothing after two rounds. 
In most situations, SWARM will not stay in the same workload state for a long time because the workload is continuously changing. Hence, the value of $\beta$ is not critical for the performance of SWARM.

\begin{figure}
\centering
\includegraphics[width=0.72\textwidth]{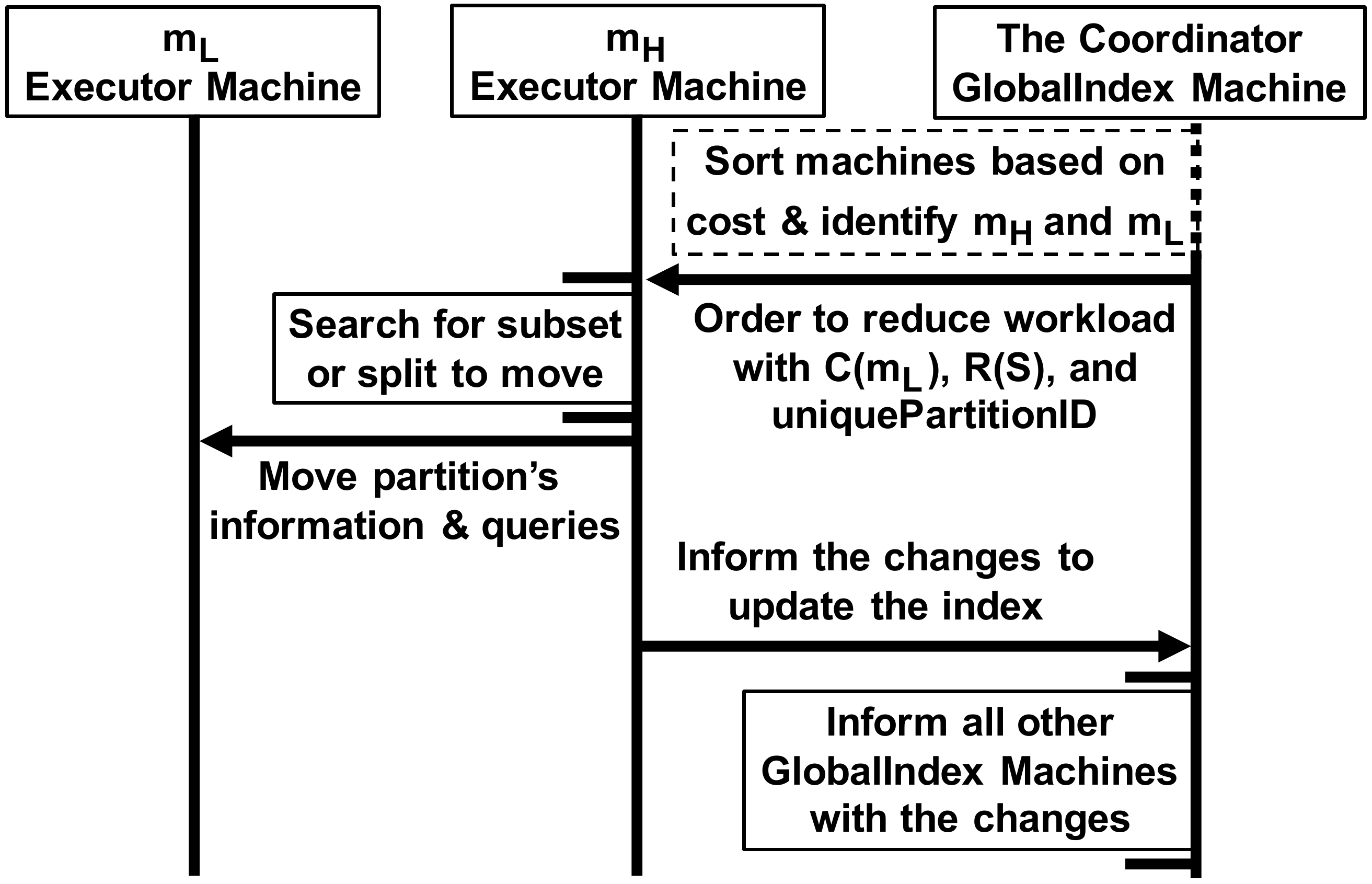}
\caption{\label{fig:balancingProcess2} Workflow of rebalancing}
\end{figure}

Figure~\ref{fig:balancingProcess2} shows the workflow when the Coordinator decides to rebalance. It sorts all executors based on their costs, and identifies the machine with the highest and lowest costs, $m_H$ and  $m_L$, respectively. The Coordinator requests from $m_H$ to reduce its workload by  migrating portions of it to $m_L$. This message contains three numbers, the cost of $m_L$ ($C(m_L)$), $R(S)$, and a unique un-used partition ID that $m_H$ can use to create new partitions, if needed. $m_H$ tries to move portions of its partitions to $m_L$, as discussed in Section~\ref{sec:workloadReduction}. If $m_H$ finds partitions to move, a new background task is created to send the partitions' information and their continuous queries to $m_L$. $m_L$ adds these partitions and continuous queries to its workload. After the move, $m_H$ reports the changes to the Coordinator that forwards the changes to all other GlobalIndex machines. They update their indexes using a latch-free background task. The cells that point to the old partition will gradually point to the new partitions that have new unique IDs. This allows the index to concurrently route new data during the update. Section~\ref{sec:integrity} discusses how the integrity of data and queries' results is preserved while migrating the workload between machines and while updating the index. If $m_H$ cannot find a feasible workload reduction, it informs the Coordinator. The latter identifies the next highest workload machine and treats it as $m_H$, and repeats the process. 

When the distribution of the workload changes, SWARM might move some old split partitions to different machines. This may result in having an executor machine controlling partitions that are adjacent, i.e., sharing a boundary. Hence, there is no benefit in keeping them separate. Moreover, they increase the overhead of maintaining separate small partitions. Adjacent partitions are combined using a background task that is triggered occasionally in all executor machines, e.g., every 5 hours. Every executor machine merges any of its partitions that form a connected rectangle. When an executor machine finds a possible merge, it creates the larger partition using a unique unused partition ID that it requests from the Coordinator. Then, the executor machine reports the changes to the Coordinator that, in turn, forwards the changes to all the other GlobalIndex machines.

\subsubsection{\textbf{Workload Reduction by Repartitioning}}\label{sec:workloadReduction}
$m_H$ can reduce its workload in one of two ways, and applies them in this order: 1) move a subset of its partitions to $m_L$. 2) split one of its partitions into two, and moves one of them to $m_L$. First, $m_H$ tries to reduce its workload using the first technique because it requires less overall overhead. If the first technique does not succeed, $m_H$ tries the second technique. The search for workload reduction is performed as a background task.

Let $P_{m_H}$ be the set of partitions that $m_H$ controls. The numerator part of the partitions' cost ($Num(C(p_i))$) is already computed, where $p_i \in P_{m_H}$. From Section~\ref{sec:balancingProcess}, $C(m_L)$, $R(S)$, and a new unique partition's ID are made available to $m_H$.
%The remaining of this section discusses the challenges of each case and purposes efficient algorithms.

%~\newline
\noindent\textbf{Searching for the Best Subset of Partitions to Move.}\hfill\\
Finding the best subset of $P_{m_H}$ to move means that after moving this subset from $m_H$ to $m_L$ both machines will have approximately equal costs and workloads. Let $C_{max}$ be the maximum cost of partitions that $m_H$ can move to $m_L$ without overloading $m_L$. Hence, $C_{max} = (C(m_H)-C(m_L))/2$. $C_{max}$ serves as a guide to ensure that the new workload plan will be better than the current one. Finding a subset of the partitions that their total costs equals $C_{max}$ will result in equal workload for $m_H$ and $m_L$ after moving the subset. $m_H$ searches for the subset that maximizes the total cost of the partitions to be moved without exceeding $C_{max}$. This is a direct application of the \textit{Subset-Sum Problem} (SSP), which is a special case of the  \textit{0-1 Knapsack Problem},
where the value of each item is equal to its weight~\cite{silvano1990knapsack}. In our case, the cost of each partition $j$ is used as the weight of each item $j$ in  SSP and $C_{max}$ is used as the capacity of the knapsack. Although SPP is NP-Hard, there is an \textit{Approximate Greedy Algorithm} that guarantees a worst-case performance ratio of $\frac{1}{2}$ with a time complexity of $O(n)$~\cite{silvano1990knapsack}. 
SWARM applies the algorithm after sorting the partitions in descending order of their costs. This increases the time complexity to $O(klog(k))$, where $k$ is the number of partitions in $m_H$. However, sorting can result in better performance on average without affecting the worst-case performance. Moreover, this extra sorting step is necessary for the splitting algorithm that might be applied if a subset is not found. Probing larger partitions first minimizes the number of moved partitions, and hence, reducing the amount of information to be sent through the network. 

\begin{algorithm}
\DontPrintSemicolon
\caption{findSubset($C_{m_L}, R_S$)}\label{alg:findSubset}
totalMoveCost = 0\;
moveSubset = empty list of partitions\;
$C_{max}=(C_{m_H}-C_{m_L})/2$\;
Sort \textit{controlledPartitionsList} based on partitions' cost from largest to lowest\;
\For {\textnormal{each partition "$p$" in \textit{controlledPartitionsList}}}{
    \If{($C_p+totalMoveCost <= C_{max}$)}{ 
		$totalMoveCost~{+}{=}~C_p$\;
		$moveSubset.add(p)$\;
		\If{($totalMoveCost == C_{max}$)}{
		    \textbf{break}\;
        }
	}
}
\Return {$moveSubset$}\;
\end{algorithm}

The procedure used by $m_H$ is presented in Algorithm~\ref{alg:findSubset}, where $C_{m_L}$ and $C_{m_H}$ are the costs of the machines with the lowest and the highest costs, respectively. $m_H$ calls this algorithm after receiving a request from  the Coordinator (with Parameters $C_{m_L}$ and $R_S$) to reduce $m_H$'s workload. $C_{m_H}$, $P_{m_H}$ ($controlledPartitionsList$), and the cost of each partition $p$ in $m_H$ ($C_p$) are available for use in the function.

%~\newline
%\vspace{-0.15in}
\noindent\textbf{Searching for Best Split for a Partition to Move.}\hfill\\
If $m_H$ fails to reduce its workload using the above technique, it tries to make $m_H$ and $m_L$ costs approximately equal using the splitting technique. $m_H$ chooses a partition $p \in P_{m_H}$ and splits it into two sub-partitions $p_1$ and $p_2$. $m_H$ calculates the expected cost difference ($C_{\textit{diff}}$) between $m_H$ and $m_L$ when $p_1$ is moved and $p_2$ is kept. $m_H$ tries to find the best split point for $p$ that will make $C_{\textit{diff}} = 0$. This is an NP-Hard problem. Thus, we use an approximate Greedy Algorithm.

SWARM considers the partition with largest cost ($p$) for splitting. If $m_H$ cannot split $p$ because it has reached the size of one cell, $m_H$ will try splitting the next largest partition in cost. The list of partitions in $m_H$ can be directly used since it has been sorted during the subset technique. $m_H$ searches for the best splitting point that results in the minimum absolute cost difference $C_{\textit{diff}}$ between $m_H$ and $m_L$. Hence, $m_H$ considers all possible vertical and horizontal split lines in $p$. The algorithm used in SWARM reduces the number of possible split lines by performing a binary search on the rows and columns of the statistics covering $p$ that contain all numbers needed to calculate $C_{\textit{diff}}$. $C_{\textit{diff}}$ is computed as follows:
\begin{equation*}
C_{\textit{diff}} = [ (C(m_H) - C(p)) + C(p_2) ] - [C(m_L) + C(p_1)]
\end{equation*}
Before the search, $m_H$ already has $C(m_H)$, $C(m_L)$, and $C(p)$. This makes the search depends only on $C(p_1)$ and $C(p_2)$. Let $sp$ be the statistic's row/column index of a split point on $p$. $C(p_1)$ and $C(p_2)$ can be computed using Eqn.~\ref{eq:c(p)} and the maintained local statistics in $p$ as follows:
\begin{align*}
C(p_1) &= N(sp) \times Q(sp) \times R(sp) / R(S)\\
Q(p_2) &= Q(p)-Q(sp)+\textit{spanQ}(sp+1)\\
R(p_2) &= R(p) - R(sp) + \textit{preSpanQ'}(sp+1)\\
C(p_2) &= [N(p)-N(sp)] \times Q(p_2) \times R(p_2) / R(S)
\end{align*}
The search ends when the algorithm finds a split point that causes $C_{\textit{diff}} = 0$. Otherwise, the search continues till the end and the split point that achieves minimum absolute value of $C_{\textit{diff}}$ is used. In the worst case, $m_H$ performs 4 binary searches: two searches on horizontal splitting points while considering the moved partition $p_1$ to be the upper or lower sub-partition, and two searches on vertical splitting points while considering $p_1$ to be the right or left sub-partition.

\section{System Integrity}\label{sec:integrity}
\subsection{Correctness During Load Balancing}
SWARM does not stop receiving and processing new data points and queries during the process of load balancing. The critical point of losing a data point or processing a data point twice can happen after identifying $m_H$ and $m_L$. $m_H$ continues to receive and process new data points and queries while searching for workload reduction. Moreover, if $m_H$ decides to split a partition, it continues to use the old partition while creating the two new sub-partitions. After $m_H$ finds either a subset of partitions or a good split, $m_H$ sends the metadata of the moved partition/s and their continuous queries to $m_L$. After the move, $m_H$ informs the Coordinator about the changes. Then, the Coordinator informs the remaining GlobalIndex machines. 
Whenever a GlobalIndex machine receives the new changes from $m_H$, it runs a latch-free background task that updates the index according to the changes while using the index for routing, as discussed in Section~\ref{sec:balancingProcess}. During the update of GlobalIndex machines, $m_H$ forwards new incoming objects that overlap the moved partition/s to $m_L$. $m_H$ keeps the metadata of the moved partition/s until their data are expired and the next load balancing round starts. Starting a new load balancing round implies that all GlobalIndex machines have finished updating their indexes, i.e., GlobalIndex machines route all new objects that overlap the moved partition/s to $m_L$. This mechanism ensures that no objects get lost or processed twice during load balancing.

\subsection{Correctness of Query Execution}
Most applications of distributed streaming systems are focused either on the current state or some limited extended state of the data. Limited extended state could be based on a time window (e.g., sliding or tumbling window) or based on data item count (count window) or some storage size (e.g., predefining the size of stored data in the window). Every distributed streaming system has a specific form of data expiration policy. Because data will eventually expire, {\bf SWARM reduces communication overhead by not moving data}. SWARM needs to know whenever data points are expired to stop tracking on which machine they are stored and to update the statistics. SWARM moves the partitions with only their continuous queries. In applications that support stateful operators (e.g., aggregate operators), the state is stored in the query not in the partition. Hence, SWARM moves the queries and their states with the migrated partition to their new executor before redirecting the stream.

As partitions are split and are moved to other machines, SWARM keeps a record in the metadata of every sub-partition that links the sub-partition to its previous responsible machine and its parent partition. Before moving a partition to $m_L$, $m_H$ adds its machine ID as the previous responsible machine and the previous $PartitionID$ as the parent partition to the metadata of the moved partition. $m_H$ continues to hold parent partitions and their data until all data become expired. Splitting or moving a partition multiple times before the data expires might lead to a chain of partitions, where each of them is linked to the previous one. Mostly, the chain of partitions will remain short because In-memory systems tend to support short windows that make data expire quickly.

A query may only need a subset of the chain of partitions to be involved in the final result. When a partition, say $p$, is to answer a query $q$, the machine responsible for $p$ will check if its parent partition, say $p_p$, exists and needs to be involved. If $p_p$ is found, its responsible machine is asked to process $q$. All involved machines send their answers directly to the machine that has $q$, say $m_q$. $m_q$ waits for the next involved machine in the chain to send the results of $q$. Every involved machine consults the next involved machine in the chain to answer $q$. Depending on whether an involved partition is expired or not, there are two ways to respond: 1)~If the partition is expired, its machine acknowledges $m_q$ and the previous involved machine in the chain that the partition is expired. Hence, the previous involved machine in the chain breaks the chain by cleaning the record of the previous responsible machine in the metadata of the partition and becoming the last machine in the chain. 2)~If the partition is not expired, $q$'s results are sent to $m_q$. Every involved machine sends an acknowledgement message to $m_q$ after it is done sending $q$ results. Acknowledgement messages contain the number of result messages that were sent and the next involved machine ID in the chain. $m_q$ keeps track of all involved machine IDs and the status of their results. $m_q$ produces the final output after receiving all result messages and acknowledgement messages from every involved machine. To produce the final output, $m_q$ combines all received answers with the answers from its partition that overlaps $q$.

For example, assume that a partition $p_1$ in executor machine $m_1$ is split into $p_2$ and $p_3$. $m_1$ adds to the metadata of both $p_2$ and $p_3$ the previous responsible machine $m_1$ and the parent partition $p_1$. $m_1$ keeps $p_2$ and moves $p_3$ to $m_2$. Now, assume that $m_2$ receives a query $q$ related to $p_3$, and it needs old data. $m_2$ finds that the previous responsible machine in the metadata of $p_3$ is $m_1$. Therefore, $m_2$ sends $q$ to $m_1$ with $p_1$'s ID as the target parent partition. If $p_1$ is not expired, $m_1$ applies $q$ on $p_1$, and sends the results to $m_2$ and an acknowledgement message. Otherwise, $m_2$ acknowledges that $p_1$ is expired to break the chain of partitions associated with $p_3$. After receiving the acknowledgement, $m_2$ produces the final results by combining $p_3$'s results with any received results.

\section{Experiments}\label{sec:experiments}
We realize SWARM in Apache Storm~\cite{toshniwal2014storm}. However, SWARM can be used with any other distributed streaming systems that process spatial data streams in tuple-at-a-time manner. As discussed in Section~\ref{sec:preliminaries}, we use a location-aware publish-subscribe application to compare SWARM  against 3 other approaches: 1)~\textit{Replicated}: New queries are replicated into all executor machines, each covering the whole space. In contrast, a new data point is sent to only one executor machine in round-robin fashion. 2)~\textit{Static Uniform Grid}: The whole space is evenly partitioned among all executor machines. 3)~\textit{Static Grid Based on History}: The partitioning of the whole space is determined based on observing a limited history of the data and query workloads. The whole space of \textit{Static Grid Based on History} is partitioned offline based on 400K data points and 200K queries taken from the dataset. SWARM's cost model is used in the third approach with the limited history to partition the workload. Hence, the costs of all executor machines are almost equal.

Experiments are performed using 6 Amazon EC2 instances. Apache Storm 1.0.0 runs in each instance over Ubuntu 18.04.2. Five instances are of type m5.2xlarge, where each instance has 8 vCPU and 32 GB of memory. Each of the five instances is divided into 8 virtual machines each having one vCPU and 4 GB of memory. This results in a total of 40 virtual machines. The last instance is of type m5.xlarge with 4 vCPU and 16 GB of memory for the Nimbus of Storm and a Zookeeper server~\cite{zookeeper}. The network bandwidth is up to 10 Gbps. 
 
Experiments are performed using a real dataset from Twitter and a synthetic query workload. The used dataset is composed of 1 Billion geotagged tweets of size 140 GB in the US. The tweets are collected from January 2014 to March 2015. The spatial data stream is made continuous and infinite by streaming the 1 Billion tweets repeatedly from the beginning each time they finish. The query workload is composed of continuous range queries. The focal points of the queries are determined using the locations of the real tweets. 

All the experiments are performed from a cold start. The used approaches store continuous queries in an R*-Tree index~\cite{beckmann1990rtree}. The grid index that divides the whole space is of size $1000 \times 1000$. This size allows small cities in the US to be covered by multiple cells. There are one million queries that are pre-loaded to every system. The spatial side lengths of queries are 0.16\% of the side length of the whole space (about the size of a university campus). As mentioned, our cluster is composed of 40 virtual machines. 10 virtual machines are Storm spouts that produce the tweets stream and the queries. The remaining 30 virtual machines are executor machines in the \textit{Replicated} approach. In SWARM and the other two static grid approaches, the 30 virtual machines are divided as 8 routing machines (GlobalIndex in SWARM) and 22 Executor machines. The machines' ratio is chosen after conducting an empirical study. We find that a good starting ratio is 1:3 (GlobalIndexes:Executors). SWARM starts a new load balancing round every 15 seconds.

\subsection{Capability and Execution Latency}
\begin{figure}
\centering
\includegraphics[trim=0in 0.05in 0in 0in, clip, width=\textwidth]{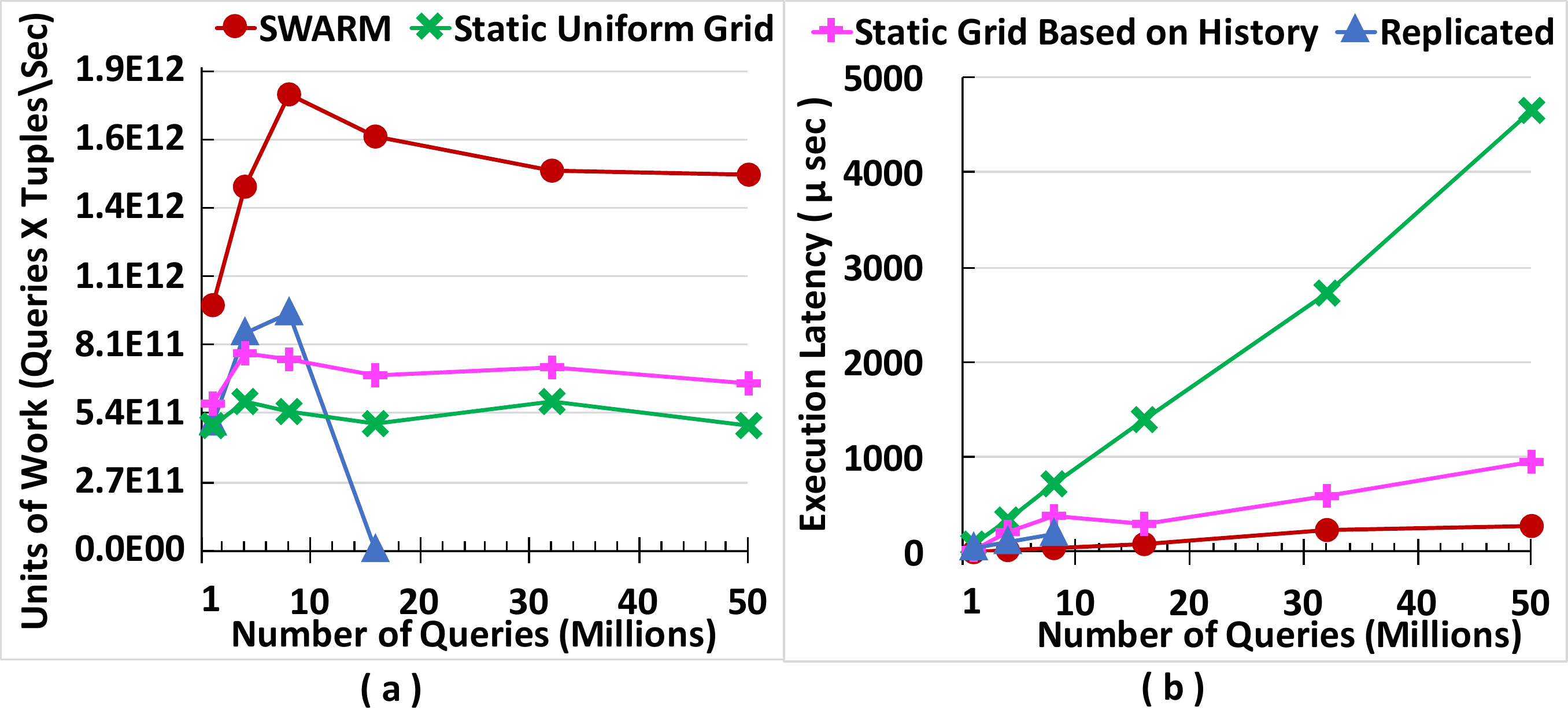}
\caption{\label{fig:PerformanceAndLatency} Capability and execution latency}
\end{figure}
One performance measure that we use is the \textit{Units of Work} measure, which is the number of tuple checks per second against queries. \textit{Units of Work} is calculated by multiplying the number of queries in the system by the number of processed tuples per second. We use \textit{Units of Work} in place of the throughput of the system as the former provides a fairer comparison because it reflects the overall amount of checks (work) that is conducted by the system regardless of the {\em selectivities} of the queries. Figure~\ref{fig:PerformanceAndLatency}a gives the average \textit{Units of Work} after running the system for an hour. SWARM outperforms the other approaches while varying the number of continuous queries in the system. On average, SWARM achieves 200\% improvement over \textit{Static Grid Based on History}. SWARM performance saturates after 32 million queries as the system reaches its peak capacity for the used tweets distribution. The \textit{Replicated} achieves better performance than both static grid approaches between 8 and 16 million queries because of better workload distribution among the system's machines. However, \textit{Replicated} fails to support more than 16 million queries due to high memory overhead as all queries and indexes are replicated on all machines.

Figure~\ref{fig:PerformanceAndLatency}b shows the average execution latency in microseconds while varying the number of queries after running the system for an hour. SWARM achieves the lowest average execution latency compared to other approaches. \textit{Replicated} achieves lower average execution latency than both static grid approaches. However, it fails to support more than 16 million queries. The incremental rate of SWARM's average execution latency is very small compared to the other approaches. SWARM reduces execution latency on average 4x compared to \textit{Static Grid Based on History}.

\subsection{Reaction to Hotspots}
Twitter real dataset contains thousands of hotspots, where lots of tweets overlap in location and in time. We are interested in observing how SWARM reacts to various types of hotspots in contrast to the approaches. Hence, several scenarios of hotspots are created by synthetically redirecting a percentage of the data spouts to a specific location in the US. The normal Twitter dataset with its hotspots is used at the beginning and at the end of the experiments' time line. By default, the locations of the data points and queries that compose a hotspot's are generated inside a square range with a spatial side length of 15\% of the whole space using a uniform distribution. All queries in a hotspot are instantiated during one minute of the hotspot's start time. This is to test how fast SWARM reacts to extreme hotspot situations. 

\begin{figure}
\centering
\includegraphics[trim=0in 0.0in 0in 0in, clip, width=\textwidth]{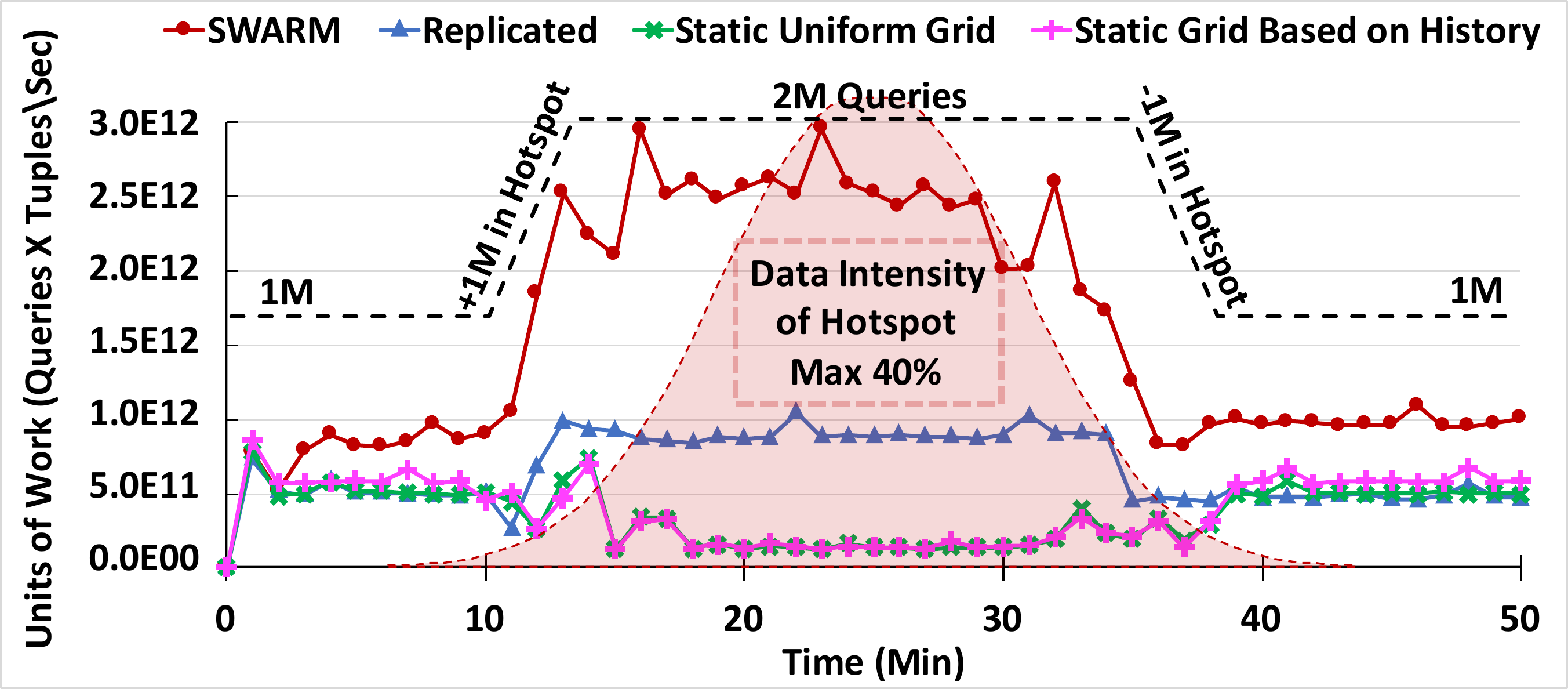}
\caption{\label{fig:OneUniformHotspot} Uniform distribution hotspot with normal distribution data intensity}
\end{figure}

\begin{figure}
\centering
\includegraphics[trim=0in 0.0in 0in 0in, clip, width=\textwidth]{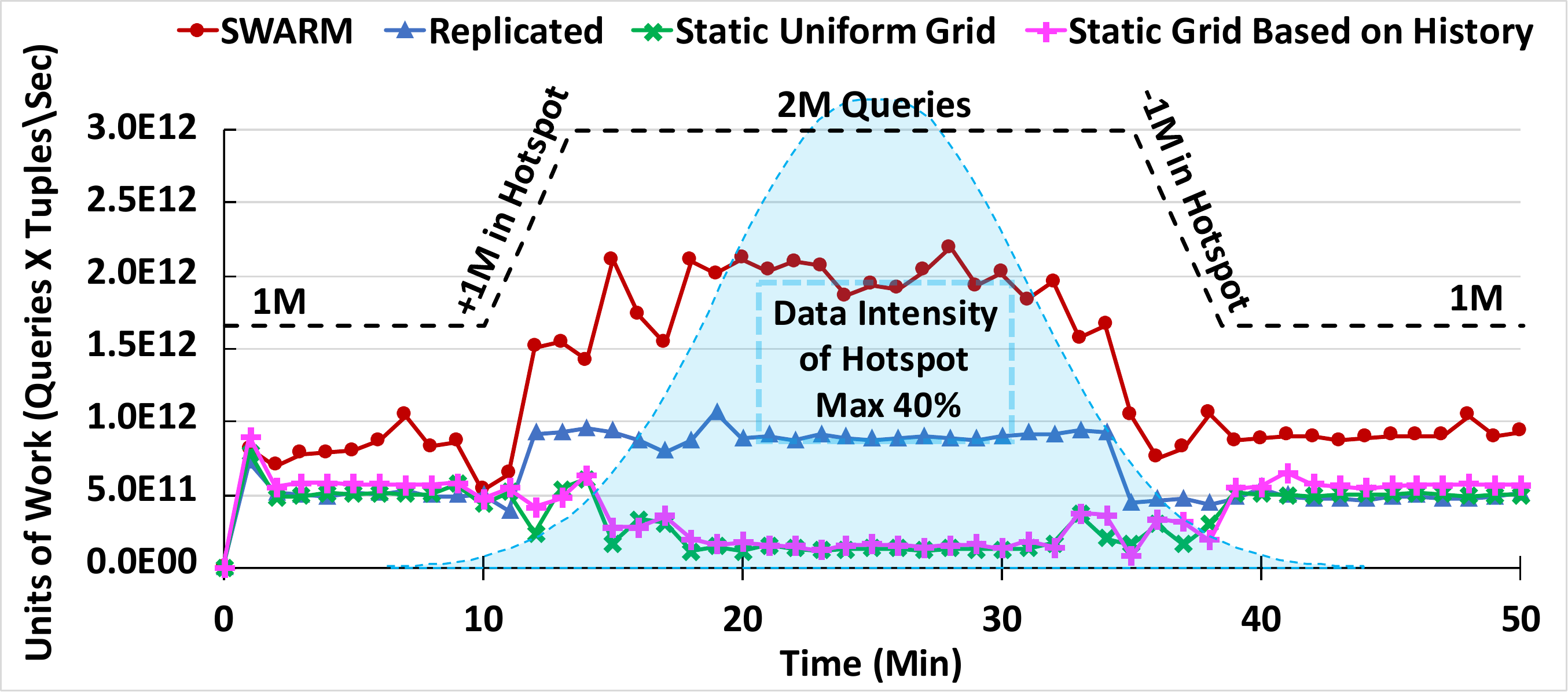}
\caption{\label{fig:OneNormalHotspot} Normal distribution hotspot with normal distribution data intensity}
\end{figure}

\begin{figure}
\centering
\includegraphics[trim=0in 0.0in 0in 0in, clip, width=0.67\textwidth]{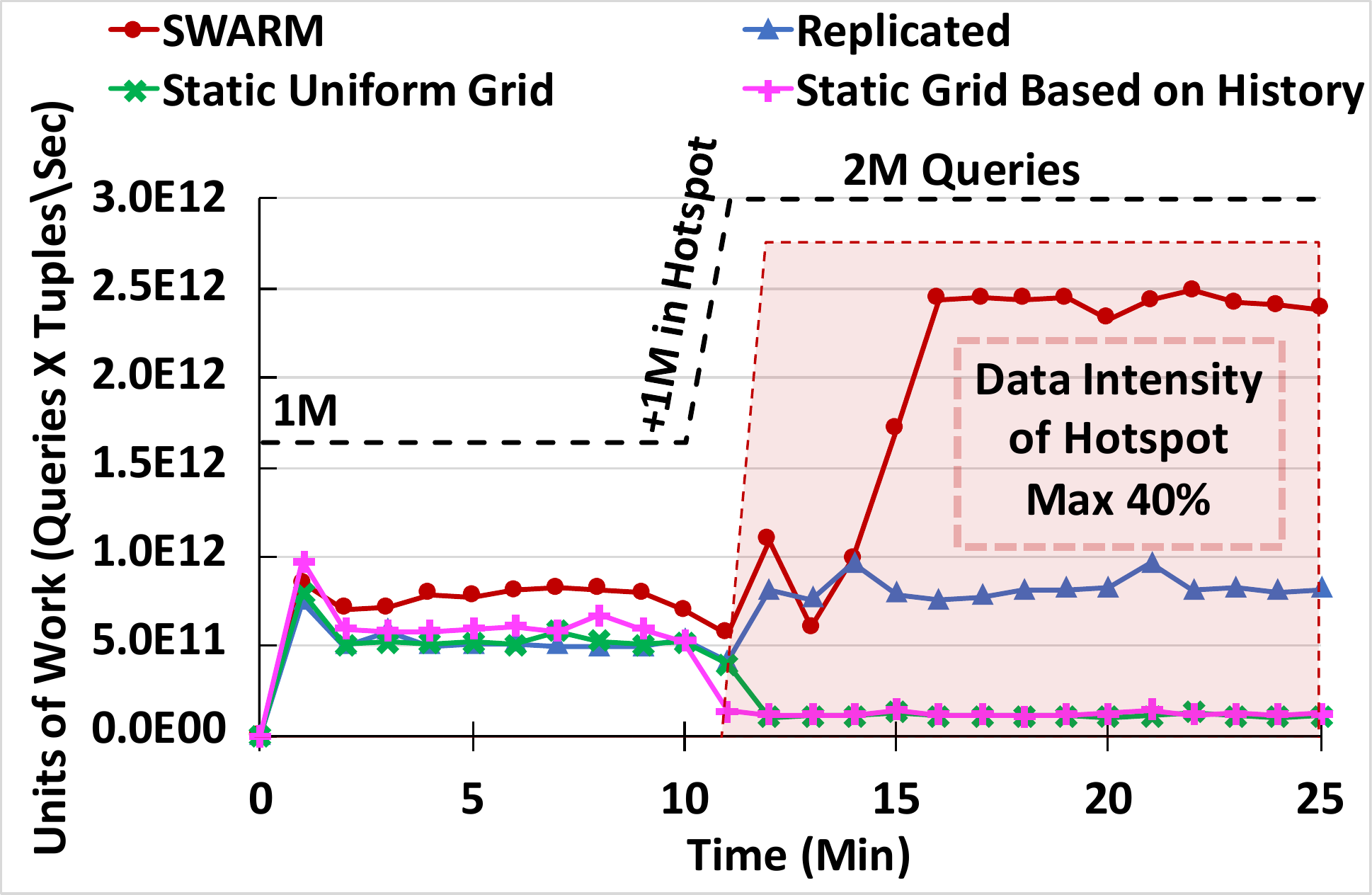}
\caption{\label{fig:OneStepFunctionHotspot} Uniform distribution hotspot with step data intensity}
\end{figure}

\begin{figure}
\centering
\includegraphics[trim=0in 0.0in 0in 0in, clip, width=\textwidth]{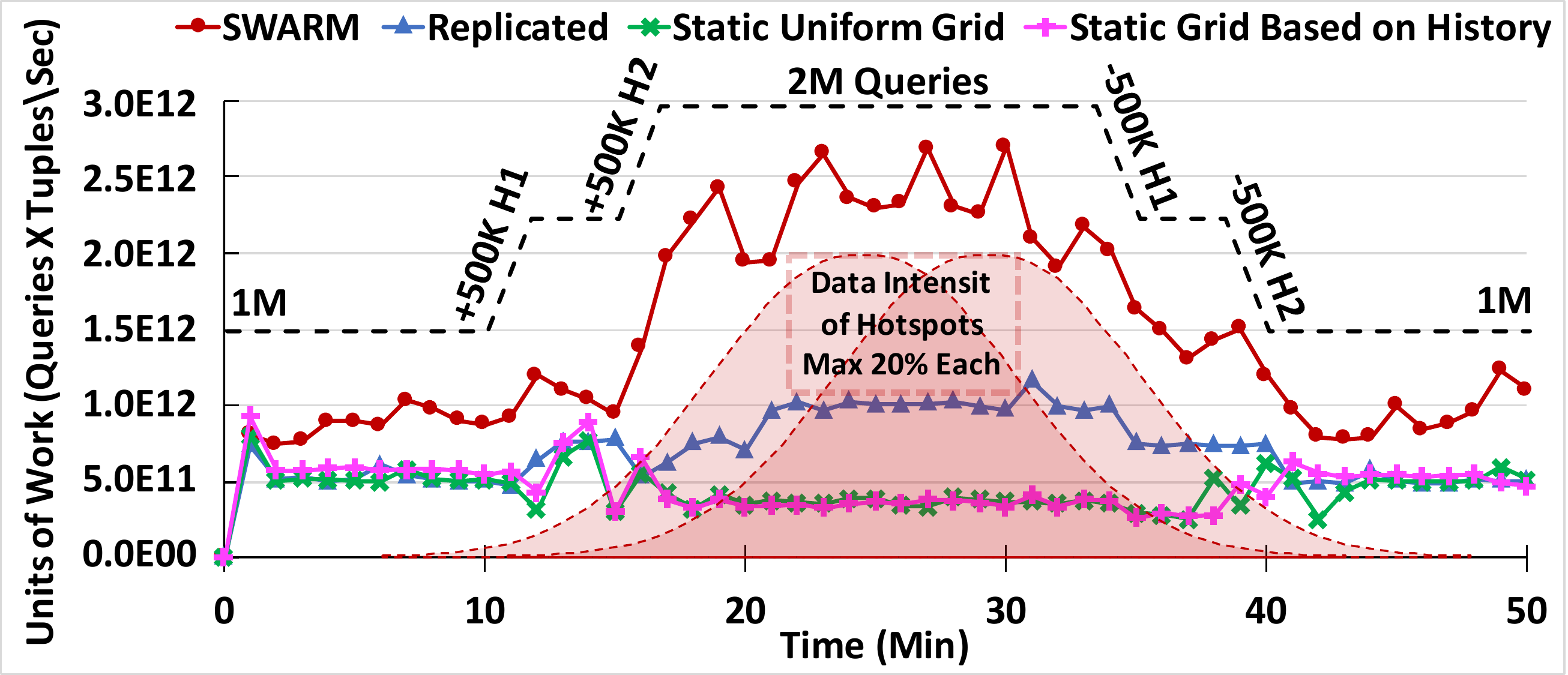}
\caption{\label{fig:TwoOverlapingHotspots} Two overlapping hotspots (H1 and H2) in different locations}
\end{figure}

\begin{figure}
\centering
\includegraphics[trim=0in 0.0in 0in 0in, clip, width=\textwidth]{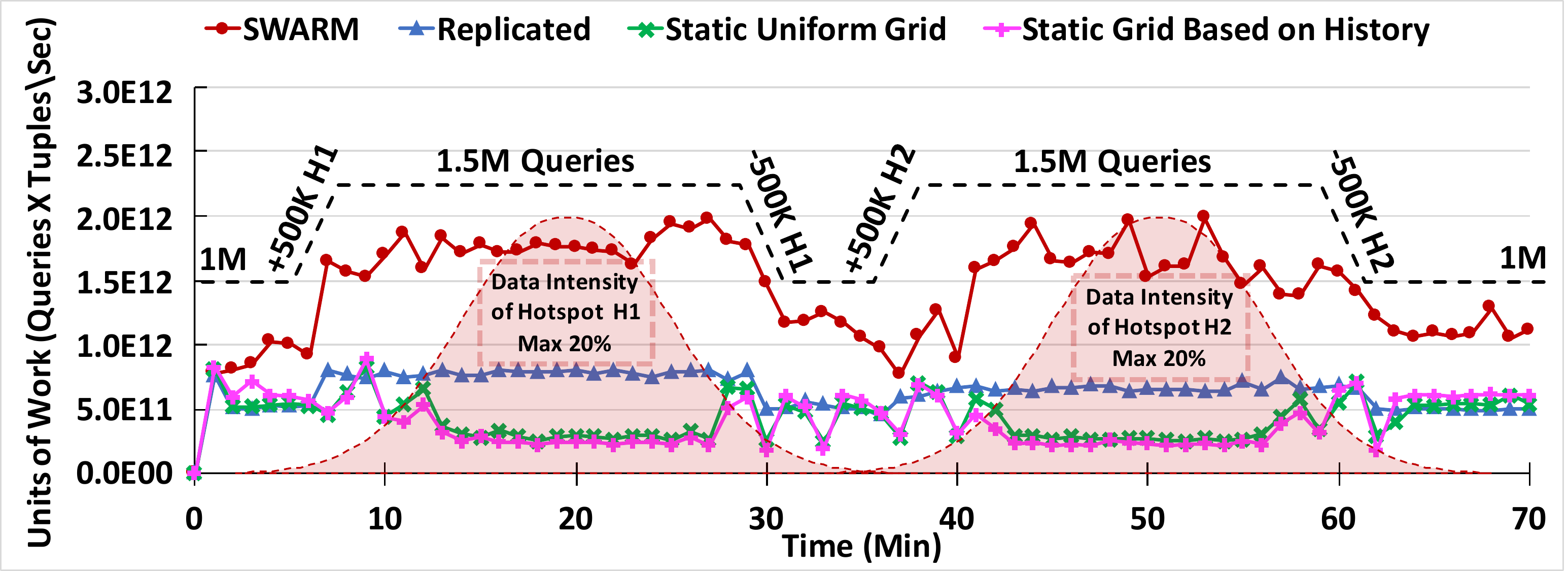}
\caption{\label{fig:TwoConsecutiveHotspots} Two consecutive hotspots (H1 and H2) in different locations}
\end{figure}

Every Figure from~\ref{fig:OneUniformHotspot} to~\ref{fig:TwoConsecutiveHotspots} shows a timeline for the average \textit{Units of Work} per minute. The axis scale of the \textit{Units of Work} is the same across all experiments. The dashed line represents the number of queries in the system. Figure~\ref{fig:OneUniformHotspot} compares the performance of all approaches with the appearance of one hotspot. The data intensity of the hotspot follows a normal distribution (the shaded area in the figure). The hotspot is created by redirecting up to 40\% of the data spouts to the lowerleft corner of the US. As the figure indicates, SWARM outperforms all other approaches. SWARM achieves higher performance during the hotspot than before and after the hotspot. SWARM's higher performance is due to having a better chance to redistribute the uniformly distributed hotspot. Both of the static grid approaches suffer during the hotspot because only a small set of their executor machines become overloaded with the hotspot. During the regular Twitter hotspots, \textit{Static Grid Based on History} achieves better performance than both \textit{Static Uniform Grid} and \textit{Replicated}. During the synthetic hotspot, it has the worst performance because its partitions are pre-determined using a limited history of Twitter's normal dataset. The sudden increase and drop in processing performance is due to the back-pressure of the spouts that periodically makes the spouts try to increase the data injection rate. The performance of all approaches return to normal after the disappearance of the hotspot. 

Figure~\ref{fig:OneNormalHotspot} gives the performance when the hotspot's data points are generated using normal distribution inside the hotspot's region instead of using a uniform distribution. The normal distribution's variance is 20\% of the hotspot's spatial side length. SWARM outperforms all the other approaches. SWARM has lower performance in the case of the normal-distribution hotspot in contrast to the uniform-distribution hotspot. The reason is that there are higher levels of spatial overlap among the data points and the queries inside the normal distribution hotspot that makes it harder for SWARM to find a new partitioning that distributes the workload evenly.

Figure~\ref{fig:OneStepFunctionHotspot} gives the performance when a uniform distribution hotspot appears directly with maximum data intensity. The data intensity of the hotspot follows a step function. Although this type of hotspots is uncommon, SWARM manages to overcome the drop in performance. SWARM experiences a sudden drop in performance immediately after the hotspot starts because SWARM does not complete redistributing the partitions while some of the machines become overloaded, and this triggers a backpressure from the Storm spouts to reduce their data injection rates. Once SWARM completes the redistribution of the partitions, the spouts' backpressure re-increases the data injection rate slowly.
\\

Figures~\ref{fig:TwoOverlapingHotspots} and~\ref{fig:TwoConsecutiveHotspots} give the results when having two concurrent hotspots in two different locations. Hotspots H1 and H2 are located in the lowerleft and upperright corners of the US, respectively. Each hotspot is created by redirecting up to 20\% of the data spouts, i.e., each hotspot has half the data intinsity of the hotspots in the previous expermints. Figure~\ref{fig:TwoOverlapingHotspots} gives the performance results when the two hotspots (H1 and H2) overlapping in time. SWARM achieves similar performance to that in Figure~\ref{fig:OneUniformHotspot}. Thus, SWARM is not affected by the number of simultaneous hotspots or their spatial locations. The intensity and distribution of the hotspots are the main factors that affect SWARM's rebalancing performance. 

Figure~\ref{fig:TwoConsecutiveHotspots} gives the performance when Hotspot H2 appears directly after Hotspot H1 disappears. This experiment illustrates that SWARM can quickly react to the changes in workload distribution. However, the performance of SWARM slightly drops at the start of H2 because SWARM needs to register the new hotspot queries as well as move some of them to other machines to rebalance the system. This slows down the processing of the new incoming data during that time. After applying the new partitioning plan, SWARM achieves similar performance as the performance during Hotspot H1.
\\

\begin{figure}
\centering
\includegraphics[trim=0in 0.05in 0in 0.05in, clip, width=0.9\textwidth]{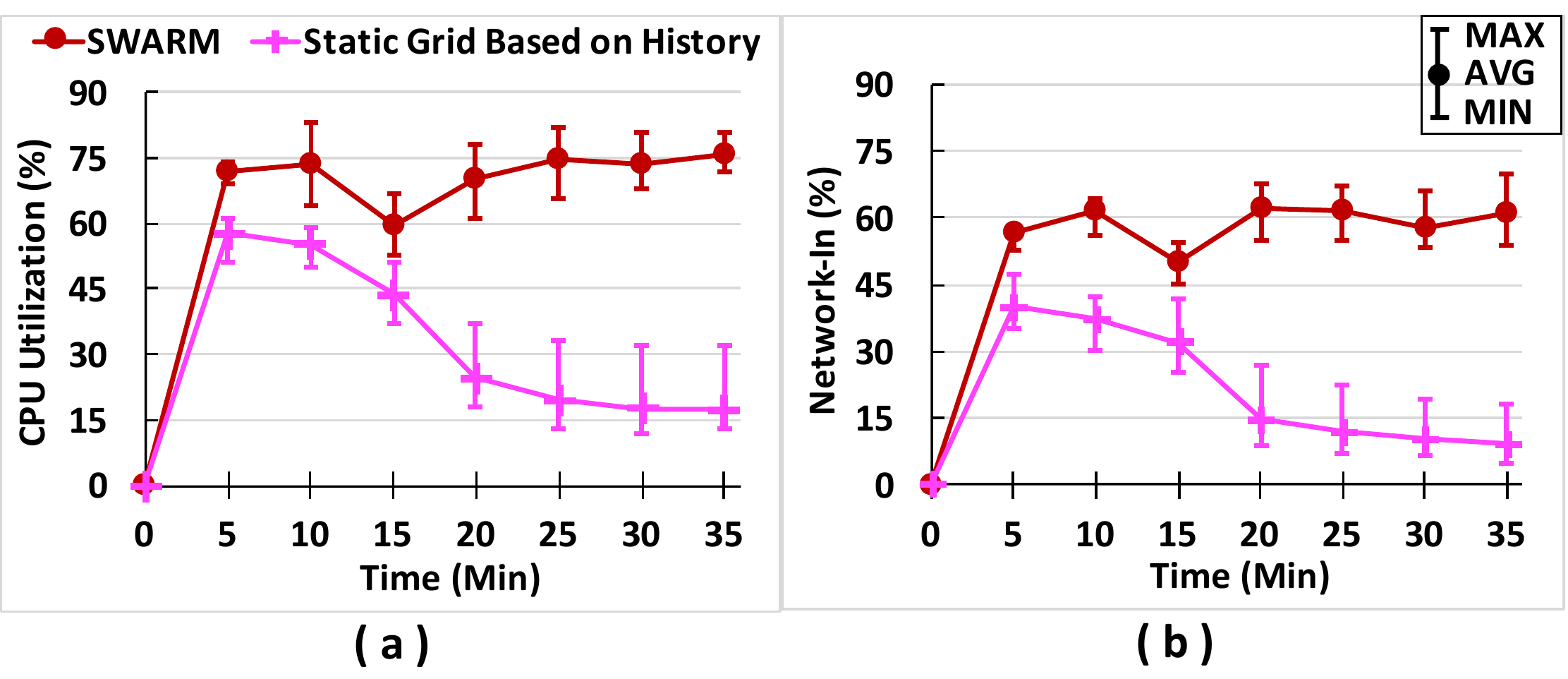}
\caption{\label{fig:CPUandNetwork} CPU and network utilization}
\end{figure}
Figure~\ref{fig:CPUandNetwork} gives the CPU and network utilization averages over 5 minutes for the cluster's virtual machines. Also, the minimum and the maximum utilization average achieved by a machine is presented. The utilization is measured during the run of the experiment presented in Figure~\ref{fig:OneUniformHotspot}. The drop in CPU and network utilization after 10 minutes of running the experiment happens because the hotspot starts at that moment. Figure~\ref{fig:CPUandNetwork}a illustrates that SWARM's average CPU utilization is improved and the gap between the minimum and maximum utilized machines gets smaller as SWARM redistributes the hotspot workload among the other underloaded machines. For \textit{Static Grid Based on History}, the overloaded machines  affect the performance of the whole system because the backpressure of the spouts makes them reduce the injection rate to match the processing rate of the slowest machine. Before running the experiment, we test the network utilization to estimate the network's achievable maximum bandwidth, which is found to be 80\% of the maximum advertised network bandwidth. Figure~\ref{fig:CPUandNetwork}b illustrates that SWARM almost reaches the highest achievable network utilization. This highlights that the bottleneck in SWARM is moved from being in the processing power to being in the network bandwidth. 

\begin{figure}
\centering
\includegraphics[trim=0in 0.0in 0in 0in, clip, width=0.8\textwidth]{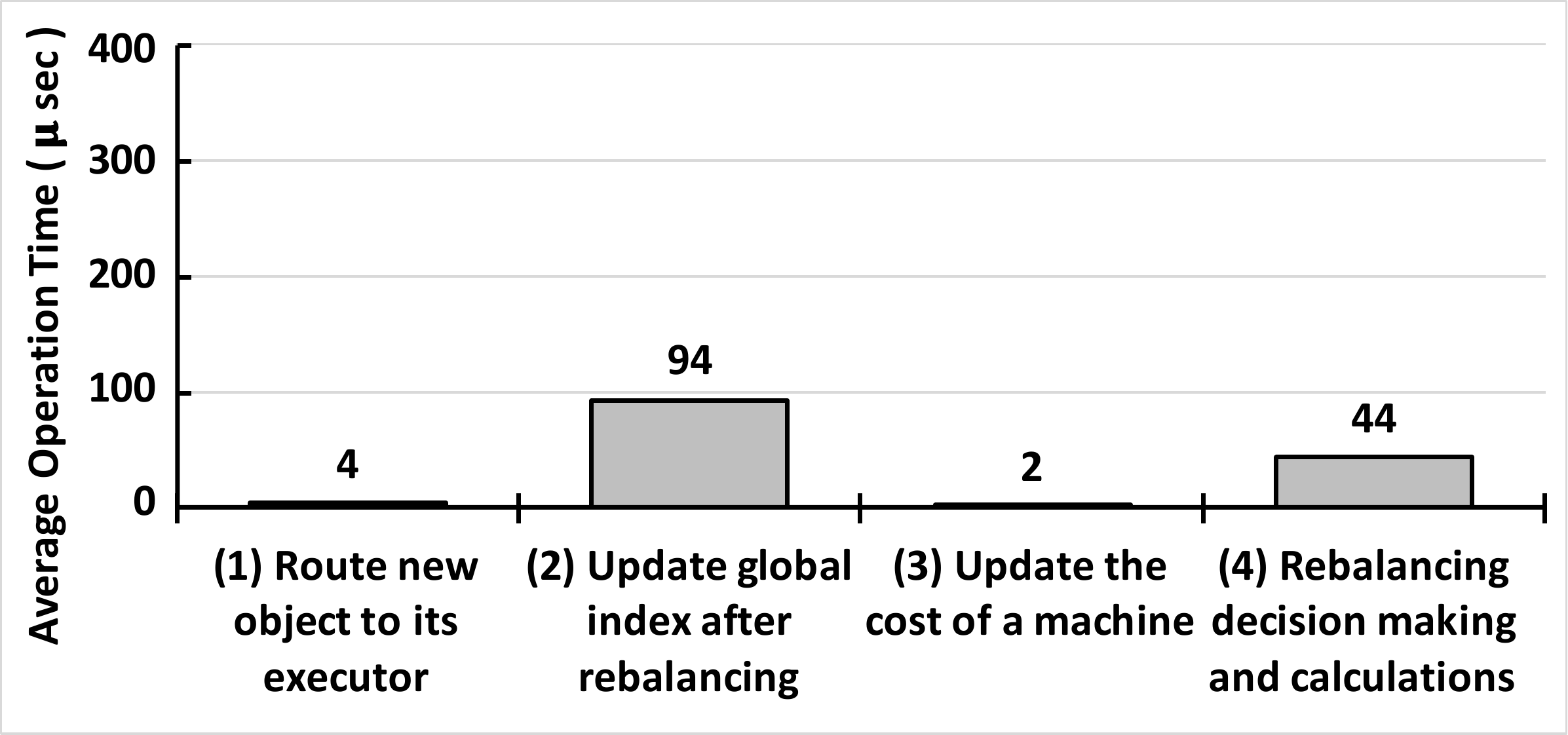}
\caption{\label{fig:OverheadGlobalIndex} Overhead of SWARM operations in GlobalIndex machines}
\end{figure}

\begin{figure}
\centering
\includegraphics[trim=0in 0.0in 0in 0in, clip, width=0.8\textwidth]{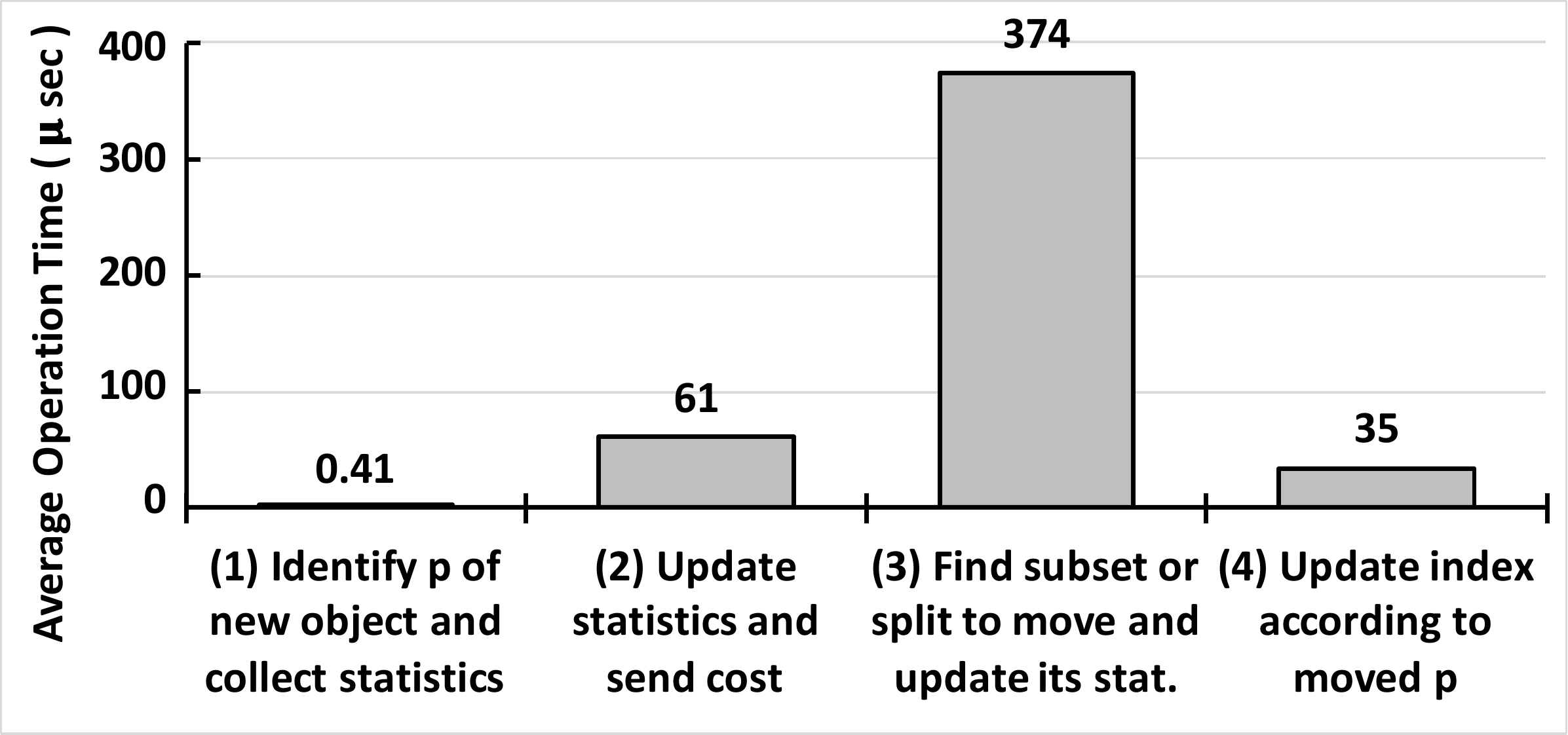}
\caption{\label{fig:OverheadExecutor} Overhead of SWARM operations in executor machines}
\end{figure}

Figures~\ref{fig:OverheadGlobalIndex} and~\ref{fig:OverheadExecutor}
illustrate the overhead of SWARM operations by showing the average time each operation takes in microseconds after running the system for an hour. Figure~\ref{fig:OverheadGlobalIndex} presents the operations of the GlobalIndex machines, while Figure~\ref{fig:OverheadExecutor} presents the operations of the executor machines. Note that as we go from operation (1) to (4) in both figures, the frequency of performing the operation and the number of machines performing it are significantly decreased. Figure~\ref{fig:OverheadGlobalIndex}-(1) gives the time it takes for the GlobalIndex machines to find the responsible executor machine for a newly received object and route this object. Figure~\ref{fig:OverheadGlobalIndex}-(2) shows the required time to update the index of a GlobalIndex machine according to a moved subset or split. Figure~\ref{fig:OverheadGlobalIndex}-(3) and~\ref{fig:OverheadGlobalIndex}-(4) shows the operations that get executed only in the Coordinator to receive the executors' cost, finding if rebalancing is needed, and identifying $m_H$ and $m_L$.

Figure~\ref{fig:OverheadExecutor}-(1) gives the time added (overhead) to the processing of a new object to identify its partition and collect its statistics. This shows the success of SWARM in minimizing the added overhead to the processing of each new object. At the end of every load-balancing round, the time of Figure~\ref{fig:OverheadExecutor}-(2) is needed to update the statistics, compute the cost, and send the cost to the Coordinator.  Figure~\ref{fig:OverheadExecutor}-(3) gives the required time for $m_H$ to find a workload reduction, update the statistics, and move the partition(s) to $m_L$. Figure~\ref{fig:OverheadExecutor}-(4) gives the required time for $m_L$ to receive a moved partition and update $m_L$'s index accordingly. 
\begin{figure}
\centering
\includegraphics[trim=0in 0.0in 0in 0in, clip, width=0.55\textwidth]{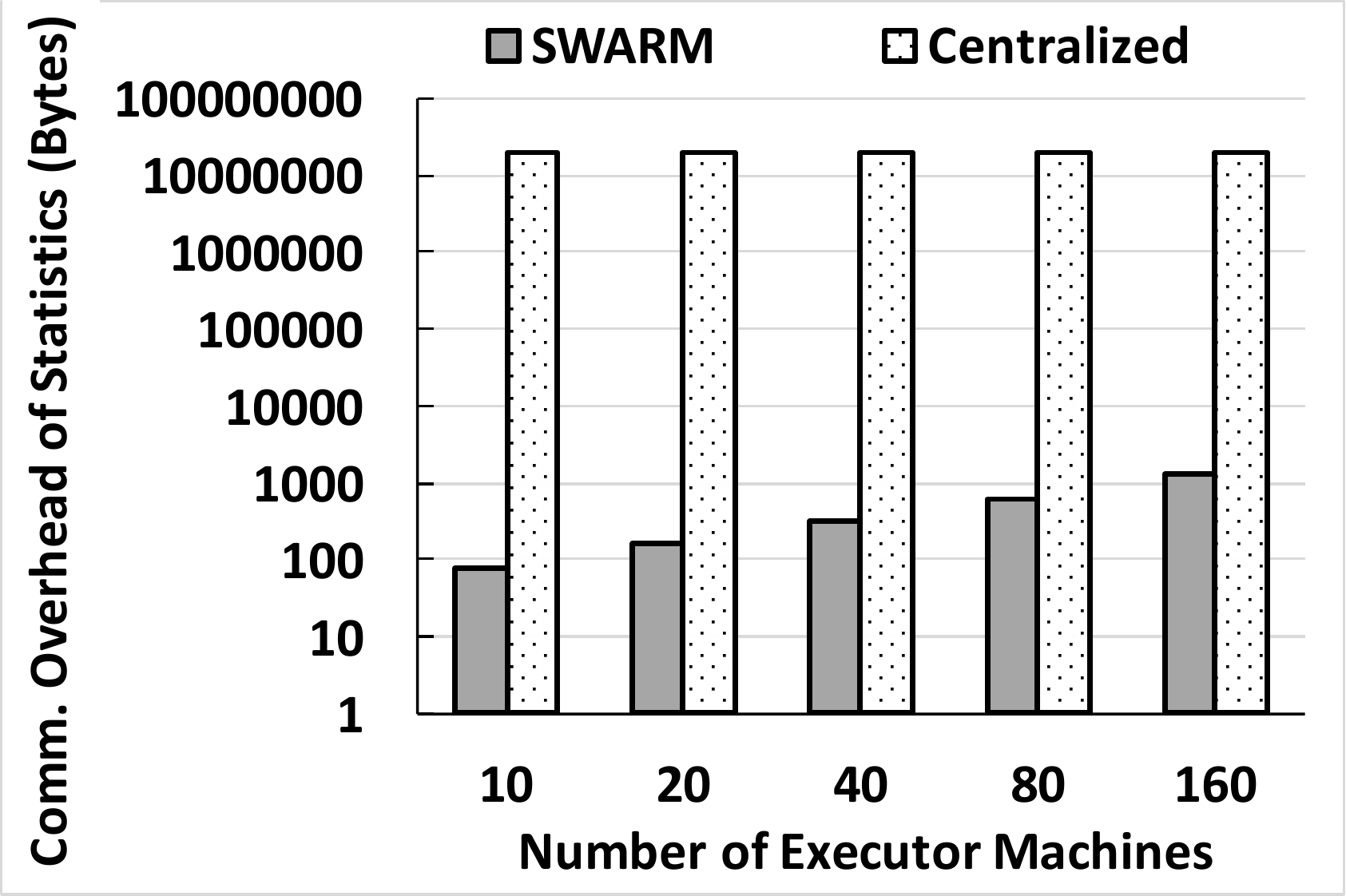}
\caption{\label{fig:OverheadNetwork} Network overhead of statistics}
\end{figure}

Figure~\ref{fig:OverheadNetwork} gives the network overhead of SWARM's decentralized statistics compared to a centralized approach. The centralized approach is following AQWA's centralized statistic technique~\cite{Aly2015aqwa}, where one number per cell is needed to count the data points, and four numbers to count the queries. The four numbers in each cell are required to use Euler Histogram~\cite{an2001selectivity,beigel1998geometry,sun2002selectivity} to count queries in a partition without re-counting queries that overlap multiple cells. Figure~\ref{fig:OverheadNetwork} compares the two approaches by measuring the number of bytes needed to be send to the Coordinator to find and apply a new partitioning plan. SWARM's decentralized approach outperforms the centralized approach because SWARM requires sending only two statistics. In contrast, the centralized approach requires sending five statistics per cell in the system, i.e., five million statistics for the $1000 \times 1000$ grid index. SWARM will always outperform the centralized approach because each machine can hold at least one cell sized partition, i.e., SWARM's machines will send 2 statistics per machine while the centralized machines will send 5. However, having every machine hold only one partition with one cell is not practical. Hence, the grid size will be increased and that will increase the amount of statistics that the centralized approach have to send. 

\section{Related Work}\label{sec:relatedWork}
In this section, we present the work related to adaptive big spatial streaming systems. We classify the work related into the following categories: (1)~centralized spatial streaming systems, (2)~General purpose big data streaming systems, and (3)~Big spatial processing systems.

%\subsection{Centralized Spatial Streaming Systems}
\textbf{Centralized spatial data streaming systems} have been developed to answer spatial queries over spatial streams, e.g.,  PLACE~\cite{mokbel2004place}, SINA~\cite{mokbel2004sina}, SEA-CNN~\cite{xiong2005sea}, and Gpac~\cite{mokbel2005gpac}. However, these systems are not scalable and cannot handle the current scale of streamed spatial data.

%\subsection{General Purpose Big Data Streaming Systems}
\begin{sloppypar}
\textbf{General-purpose big data systems} provide an infrastructure to scale up the batch and real-time processing. General-purpose big data systems are either batch-oriented or stream-oriented. Examples of batch-oriented include Hadoop~\cite{hadoop} and Spark~\cite{spark}. Batch-oriented systems require minutes or even hours to process data and are not suited for real-time processing. Yahoo S4~\cite{neumeyer2010s4}, Apache Samza~\cite{smaza}, Apache Storm~\cite{toshniwal2014storm}, Twitter Heron~\cite{kulkarni2015twitter}, and Spark Streaming~\cite{zaharia2012discretized} are examples of stream-oriented systems that can process data in real-time with latencies ranging between milliseconds up to few seconds. However, these systems are not optimized for spatial data processing and are not adaptive.
\end{sloppypar}

%\subsection{Big Spatial Data Management Systems}
\begin{sloppypar}
To enable the scalable processing of big spatial data, several general-purpose big data systems have been extended with spatial indexing and querying techniques. HadoopGIS~\cite{aji2013hadoop}, SATO~\cite{vo2014sato}, and SpatialHadoop~\cite{eldawy2015spatialhadoop} are \noindent\textbf{big spatial processing systems} on top of Hadoop. LocationSpark~\cite{tang2016locationspark}, Cruncher~\cite{abdelhamid2016cruncher}, Simba~\cite{xie2016simba}, SparkGIS~\cite{baig2017sparkgis} are spatial extensions to Spark. All these systems do not offer real-time big spatial data processing. Most of the existing big spatial data streaming systems use static data partitioning schemes to distribute the workload among machines. Zhang et al.~\cite{zhang2016real} extends Storm with static spatial partitioning to enable real-time spatial data processing. However, these techniques are not effective in spatial applications due to the rapid changes in data and query distributions. SWARM enables adaptive spatial processing over any distributed streaming system that works in a pipeline fashion including Storm.
\end{sloppypar}

%\subsection{Adaptive Data Management Systems}
\begin{sloppypar}
Several \textbf{adaptive batch and streaming management systems} have recently been proposed to handle any variabilities in the underlying workload.PKG2 and PKG5~\cite{nasir2015power,nasir2016two} are stream partitioning schemes that evenly distribute the received workload for each key among a limited number of the system's machines. PKG is not built for spatial applications. Therefore, SWARM has the leverage to change the spatial boundaries of partitions to distribute the workload of a hotspot. Moreover, SWARM is not forced to distribute the workload of a hotspot over a specific number of machines. SWARM can distribute the workload among all executor machines, if necessary. Amoeba~\cite{shanbhag2017robust,shanbhag2016moeba} is an adaptive data partitioning scheme in relational systems. Amoeba does not consider real-time stream processing. AQWA~\cite{Aly2015aqwa} is an adaptive spatial processing system on top of Hadoop. AQWA distributes new batches of data into HDFS files offline before starting to process the queries. Its distribution is based on the collected statistics in one master node. This cannot work for processing spatial streams in real-time. Cruncher~\cite{abdelhamid2016cruncher} is a proposal for adaptive spatial stream processing on top of Spark. However, Cruncher works only on micro-batch stream processing that has relatively high latency, i.e., seconds. However, SWARM is able to adaptively process spatial data in real-time with minimal latency. Tornado~\cite{mahmood2018adaptive,mahmood2015tornado} and PS2Stream~\cite{chen2017distributed} are adaptive spatio-textual streaming systems that are based on the Storm streaming system. These systems handle spatio-textual data while SWARM focuses on spatial processing.
\end{sloppypar}

\section{Conclusions}\label{sec:conclusions}
This paper introduces SWARM, a light-weight adaptivity protocol that continuously monitors the workload across the distributed machines of spatial data streaming systems. SWARM adjusts the workload distribution as soon as performance bottlenecks get detected. SWARM is able to handle multiple query-execution and data-persistence models. SWARM requires minimal changes to the original code of applications. A probabilistic cost model is introduced to help find the best partitioning plan. SWARM introduces a new statistics structure that requires minimal overhead. Greedy algorithms are presented to efficiently find the best subset or split of partition/s. SWARM preserves the system integrity while repartitioning without halting the system. SWARM is tested and is compared against other static approaches using an application that processes a real dataset from Twitter. On average, SWARM achieves 200\% improvement over a static grid approach that is partitioned based on a limited history of the workload. Moreover, SWARM reduces execution latency on average 4x compared with the other approaches.

\section*{Acknowledgements}
Walid G. Aref acknowledges the support of the National Science Foundation under Grant Numbers: IIS-1910216 and III-1815796.
% ensure same length columns on last page (might need two sub-sequent latex runs)
\balance

%%
%% The next two lines define the bibliography style to be used, and
%% the bibliography file.
\bibliographystyle{plain}
\bibliography{mainFile}

\begin{thebibliography}{10}

\bibitem{hadoop}
Apatche {H}adoop.
\newblock \url{http://hadoop.apache.org/}, 2020.

\bibitem{smaza}
Apatche {S}amza.
\newblock \url{http://samza.apache.org/}, 2020.

\bibitem{spark}
Apatche {S}ark.
\newblock \url{http://spark.apache.org/}, 2020.

\bibitem{zookeeper}
Apatche {Z}ookeeper.
\newblock \url{https://zookeeper.apache.org}, 2020.

\bibitem{internetStats}
Internet live stats.
\newblock \url{https://internetlivestats.com/}, 2020.

\bibitem{abdelhamid2016cruncher}
Ahmed~S Abdelhamid, Mingjie Tang, Ahmed~M Aly, Ahmed~R Mahmood, Thamir Qadah,
  Walid~G Aref, and Saleh Basalamah.
\newblock Cruncher: Distributed in-memory processing for location-based
  services.
\newblock In {\em Data Engineering (ICDE), 2016 IEEE 32nd International
  Conference on}, pages 1406--1409. IEEE, 2016.

\bibitem{aji2013hadoop}
Ablimit Aji, Fusheng Wang, Hoang Vo, Rubao Lee, Qiaoling Liu, Xiaodong Zhang,
  and Joel Saltz.
\newblock Hadoop gis: a high performance spatial data warehousing system over
  mapreduce.
\newblock {\em Proceedings of the VLDB Endowment}, 6(11):1009--1020, 2013.

\bibitem{Aly2016kangaroo}
Ahmed~M. Aly, Hazem Elmeleegy, Yan Qi, and Walid Aref.
\newblock Kangaroo: Workload-aware processing of range data and range queries
  in hadoop.
\newblock In {\em Proceedings of the Ninth ACM International Conference on Web
  Search and Data Mining}, WSDM '16, pages 397--406, New York, NY, USA, 2016.
  ACM.

\bibitem{Aly2015aqwa}
Ahmed~M. Aly, Ahmed~R. Mahmood, Mohamed~S. Hassan, Walid~G. Aref, Mourad
  Ouzzani, Hazem Elmeleegy, and Thamir Qadah.
\newblock Aqwa: Adaptive query workload aware partitioning of big spatial data.
\newblock {\em Proc. VLDB Endow.}, 8(13):2062--2073, September 2015.

\bibitem{an2001selectivity}
Ning An, Zhen-Yu Yang, and Anand Sivasubramaniam.
\newblock Selectivity estimation for spatial joins.
\newblock In {\em Proceedings 17th International Conference on Data
  Engineering}, pages 368--375. IEEE, 2001.

\bibitem{baig2017sparkgis}
Furqan Baig, Hoang Vo, Tahsin Kurc, Joel Saltz, and Fusheng Wang.
\newblock Sparkgis: Resource aware efficient in-memory spatial query
  processing.
\newblock In {\em Proceedings of the 25th ACM SIGSPATIAL International
  Conference on Advances in Geographic Information Systems}, page~28. ACM,
  2017.

\bibitem{beckmann1990rtree}
Norbert Beckmann, Hans-Peter Kriegel, Ralf Schneider, and Bernhard Seeger.
\newblock The r*-tree: An efficient and robust access method for points and
  rectangles.
\newblock In {\em Proceedings of the 1990 ACM SIGMOD International Conference
  on Management of Data}, SIGMOD '90, pages 322--331, New York, NY, USA, 1990.
  ACM.

\bibitem{beigel1998geometry}
Richard Beigel and Egemen Tanin.
\newblock The geometry of browsing.
\newblock In {\em Latin American Symposium on Theoretical Informatics}, pages
  331--340. Springer, 1998.

\bibitem{chen2017distributed}
Zhida Chen, Gao Cong, Zhenjie Zhang, Tom~ZJ Fuz, and Lisi Chen.
\newblock Distributed publish/subscribe query processing on the spatio-textual
  data stream.
\newblock In {\em Data Engineering (ICDE), 2017 IEEE 33rd International
  Conference on}, pages 1095--1106. IEEE, 2017.

\bibitem{choi2015processing}
Dojin Choi, Seokil Song, Bosung Kim, and Insu Bae.
\newblock Processing moving objects and traffic events based on spark
  streaming.
\newblock In {\em Disaster Recovery and Business Continuity (DRBC), 2015 8th
  International Conference on}, pages 4--7. IEEE, 2015.

\bibitem{eldawy2015spatialhadoop}
Ahmed Eldawy and Mohamed~F Mokbel.
\newblock Spatialhadoop: A mapreduce framework for spatial data.
\newblock In {\em Data Engineering (ICDE), 2015 IEEE 31st International
  Conference on}, pages 1352--1363. IEEE, 2015.

\bibitem{herlihy1990methodology}
Maurice Herlihy.
\newblock A methodology for implementing highly concurrent data structures.
\newblock In {\em ACM SIGPLAN Notices}, volume~25, pages 197--206. ACM, 1990.

\bibitem{herlihy1991wait}
Maurice Herlihy.
\newblock Wait-free synchronization.
\newblock {\em ACM Transactions on Programming Languages and Systems (TOPLAS)},
  13(1):124--149, 1991.

\bibitem{herlihy2003obstruction}
Maurice Herlihy, Victor Luchangco, and Mark Moir.
\newblock Obstruction-free synchronization: Double-ended queues as an example.
\newblock In {\em 23rd International Conference on Distributed Computing
  Systems, 2003. Proceedings.}, pages 522--529. IEEE, 2003.

\bibitem{horikawa2013latch}
Takashi Horikawa.
\newblock Latch-free data structures for dbms: design, implementation, and
  evaluation.
\newblock In {\em Proceedings of the 2013 ACM SIGMOD International Conference
  on Management of Data}, pages 409--420. ACM, 2013.

\bibitem{kulkarni2015twitter}
Sanjeev Kulkarni, Nikunj Bhagat, Maosong Fu, Vikas Kedigehalli, Christopher
  Kellogg, Sailesh Mittal, Jignesh~M Patel, Karthik Ramasamy, and Siddarth
  Taneja.
\newblock Twitter heron: Stream processing at scale.
\newblock In {\em Proceedings of the 2015 ACM SIGMOD International Conference
  on Management of Data}, pages 239--250. ACM, 2015.

\bibitem{lee2015distributed}
Yunsou Lee and Seokil Song.
\newblock Distributed indexing methods for moving objects based on spark
  stream.
\newblock {\em International Journal of Contents}, 11(1):69--72, 2015.

\bibitem{mahmood2015tornado}
Ahmed~R Mahmood, Ahmed~M Aly, Thamir Qadah, El~Kindi Rezig, Anas Daghistani,
  Amgad Madkour, Ahmed~S Abdelhamid, Mohamed~S Hassan, Walid~G Aref, and Saleh
  Basalamah.
\newblock Tornado: A distributed spatio-textual stream processing system.
\newblock {\em PVLDB}, 8(12):2020--2023, 2015.

\bibitem{mahmood2018adaptive}
Ahmed~R Mahmood, Anas Daghistani, Ahmed~M Aly, Mingjie Tang, Saleh Basalamah,
  Sunil Prabhakar, and Walid~G Aref.
\newblock Adaptive processing of spatial-keyword data over a distributed
  streaming cluster.
\newblock In {\em Proceedings of the 26th ACM SIGSPATIAL International
  Conference on Advances in Geographic Information Systems}, pages 219--228.
  ACM, 2018.

\bibitem{mokbel2016thinkingSpatial}
Mohamed~F Mokbel.
\newblock Thinking spatial, {ACM SIGMOD B}log.
\newblock \url{http://wp.sigmod.org/?p=2012}, 2016.

\bibitem{mokbel2005gpac}
Mohamed~F Mokbel and Walid~G Aref.
\newblock Gpac: generic and progressive processing of mobile queries over
  mobile data.
\newblock In {\em Proceedings of the 6th international conference on Mobile
  data management}, pages 155--163. ACM, 2005.

\bibitem{mokbel2004sina}
Mohamed~F Mokbel, Xiaopeing Xiong, and Walid~G Aref.
\newblock Sina: Scalable incremental processing of continuous queries in
  spatio-temporal databases.
\newblock In {\em Proceedings of the 2004 ACM SIGMOD international conference
  on Management of data}, pages 623--634. ACM, 2004.

\bibitem{mokbel2004place}
Mohamed~F Mokbel, Xiaopeng Xiong, Walid~G Aref, Susanne~E Hambrusch, Sunil
  Prabhakar, and Moustafa~A Hammad.
\newblock Place: a query processor for handling real-time spatio-temporal data
  streams.
\newblock In {\em Proceedings of the Thirtieth international conference on Very
  large data bases-Volume 30}, pages 1377--1380. VLDB Endowment, 2004.

\bibitem{nasir2015power}
Muhammad Anis~Uddin Nasir, Gianmarco De~Francisci Morales, David
  Garcia-Soriano, Nicolas Kourtellis, and Marco Serafini.
\newblock The power of both choices: Practical load balancing for distributed
  stream processing engines.
\newblock In {\em 2015 IEEE 31st International Conference on Data Engineering},
  pages 137--148. IEEE, 2015.

\bibitem{nasir2016two}
Muhammad Anis~Uddin Nasir, Gianmarco De~Francisci Morales, Nicolas Kourtellis,
  and Marco Serafini.
\newblock When two choices are not enough: Balancing at scale in distributed
  stream processing.
\newblock In {\em 2016 IEEE 32nd International Conference on Data Engineering
  (ICDE)}, pages 589--600. IEEE, 2016.

\bibitem{neumeyer2010s4}
Leonardo Neumeyer, Bruce Robbins, Anish Nair, and Anand Kesari.
\newblock S4: Distributed stream computing platform.
\newblock In {\em Data Mining Workshops (ICDMW), 2010 IEEE International
  Conference on}, pages 170--177. IEEE, 2010.

\bibitem{shanbhag2016moeba}
Anil Shanbhag, Alekh Jindal, Yi~Lu, and Samuel Madden.
\newblock A moeba: a shape changing storage system for big data.
\newblock {\em Proceedings of the VLDB Endowment}, 9(13):1569--1572, 2016.

\bibitem{shanbhag2017robust}
Anil Shanbhag, Alekh Jindal, Samuel Madden, Jorge Quiane, and Aaron~J Elmore.
\newblock A robust partitioning scheme for ad-hoc query workloads.
\newblock In {\em Proceedings of the 2017 Symposium on Cloud Computing}, pages
  229--241. ACM, 2017.

\bibitem{silvano1990knapsack}
Martello Silvano and Toth Paolo.
\newblock Knapsack problems: algorithms and computer implementations, 1990.

\bibitem{song2016parallel}
Ge~Song.
\newblock {\em Parallel and continuous join processing for data stream}.
\newblock PhD thesis, Universit{\'e} Paris-Saclay, 2016.

\bibitem{sun2002selectivity}
Chengyu Sun, Divyakant Agrawal, and Amr El~Abbadi.
\newblock Selectivity estimation for spatial joins with geometric selections.
\newblock In {\em International Conference on Extending Database Technology},
  pages 609--626. Springer, 2002.

\bibitem{tang2016locationspark}
Mingjie Tang, Yongyang Yu, Qutaibah~M Malluhi, Mourad Ouzzani, and Walid~G
  Aref.
\newblock Locationspark: a distributed in-memory data management system for big
  spatial data.
\newblock {\em Proceedings of the VLDB Endowment}, 9(13):1565--1568, 2016.

\bibitem{toshniwal2014storm}
Ankit Toshniwal, Siddarth Taneja, Amit Shukla, Karthik Ramasamy, Jignesh~M
  Patel, Sanjeev Kulkarni, Jason Jackson, Krishna Gade, Maosong Fu, Jake
  Donham, et~al.
\newblock Storm@ twitter.
\newblock In {\em Proceedings of the 2014 ACM SIGMOD international conference
  on Management of data}, pages 147--156. ACM, 2014.

\bibitem{vo2014sato}
Hoang Vo, Ablimit Aji, and Fusheng Wang.
\newblock Sato: a spatial data partitioning framework for scalable query
  processing.
\newblock In {\em Proceedings of the 22nd ACM SIGSPATIAL International
  Conference on Advances in Geographic Information Systems}, pages 545--548.
  ACM, 2014.

\bibitem{wu2012parallelizing}
Sai Wu, Vibhore Kumar, Kun-Lung Wu, and Beng~Chin Ooi.
\newblock Parallelizing stateful operators in a distributed stream processing
  system: how, should you and how much?
\newblock In {\em Proceedings of the 6th ACM International Conference on
  Distributed Event-Based Systems}, pages 278--289. ACM, 2012.

\bibitem{xie2016simba}
Dong Xie, Feifei Li, Bin Yao, Gefei Li, Liang Zhou, and Minyi Guo.
\newblock Simba: Efficient in-memory spatial analytics.
\newblock In {\em Proceedings of the 2016 International Conference on
  Management of Data}, pages 1071--1085. ACM, 2016.

\bibitem{xiong2005sea}
Xiaopeng Xiong, Mohamed~F Mokbel, and Walid~G Aref.
\newblock Sea-cnn: Scalable processing of continuous k-nearest neighbor queries
  in spatio-temporal databases.
\newblock In {\em Data Engineering, 2005. ICDE 2005. Proceedings. 21st
  International Conference on}, pages 643--654. IEEE, 2005.

\bibitem{yu2015scalable}
Ziqiang Yu, Yang Liu, Xiaohui Yu, and Ken~Q Pu.
\newblock Scalable distributed processing of k nearest neighbor queries over
  moving objects.
\newblock {\em IEEE Transactions on Knowledge and Data Engineering},
  27(5):1383--1396, 2015.

\bibitem{zaharia2012discretized}
Matei Zaharia, Tathagata Das, Haoyuan Li, Scott Shenker, and Ion Stoica.
\newblock Discretized streams: An efficient and fault-tolerant model for stream
  processing on large clusters.
\newblock {\em HotCloud}, 12:10--10, 2012.

\bibitem{zhang2016real}
Feng Zhang, Ye~Zheng, Dengping Xu, Zhenhong Du, Yingzhi Wang, Renyi Liu, and
  Xinyue Ye.
\newblock Real-time spatial queries for moving objects using storm topology.
\newblock {\em ISPRS International Journal of Geo-Information}, 5(10):178,
  2016.

\end{thebibliography}

\end{document}